\documentclass[prd,twocolumn,nofootinbib,showpacs,amsmath,amssymb,floatfix,eqsecnum]{revtex4-1}

\usepackage{graphicx,color,framed}
\usepackage{hyperref}
\usepackage{times}
\usepackage{enumerate}
\usepackage{lipsum}

\hypersetup{
    colorlinks=true, 
    linktoc=all,     
    linkcolor=blue,  
}

\def \beq {\begin{equation}}
\def \eeq {\end{equation}}
\def \beqa {\begin{eqnarray}}
\def \eeqa {\end{eqnarray}}
\def \bseq {\begin{subequations}}
\def \eseq {\end{subequations}}

\newcommand \al {\alpha}

\newcommand \un {\underline}
\newcommand \ep {\epsilon}
\newcommand \lam {\lambda}
\newcommand \pd {\partial}

\newcommand \mb {\mathbf}
\newcommand \mbs {\boldsymbol}

\newcommand \nnb {\nonumber}

\newcommand \vphi {\varphi}
\newcommand \vth {\vartheta}

\newcommand \A {\mathcal{A}}
\newcommand \B {\mathcal{B}}
\newcommand \M {\mathcal{M}}

\begin{document}

\title{Perturbative and global anomalies in bosonic analogues of integer quantum Hall and topological insulator phases}

\author{Matthew F. Lapa}
\email[email address: ]{lapa2@illinois.edu}
\affiliation{Department of Physics and Institute for Condensed Matter Theory, University of Illinois at Urbana-Champaign, 
Champaign, Illinois 61801-3080, USA}
\author{Taylor L. Hughes}
\affiliation{Department of Physics and Institute for Condensed Matter Theory, University of Illinois at Urbana-Champaign, 
Champaign, Illinois 61801-3080, USA}


\begin{abstract}

We study perturbative and global anomalies at the boundaries of bosonic analogues of integer quantum Hall (BIQH) and topological 
insulator (BTI) phases using a description of the boundaries of these phases in terms of a nonlinear sigma model (NLSM) with Wess-Zumino term. One of the main results of the paper is that these anomalies are robust against arbitrary smooth deformations of the 
target space of the NLSM which describes the phase, provided that the deformations also respect the symmetry of the phase.
In the first part of the paper we discuss the perturbative $U(1)$ anomaly at the boundary of BIQH states in all odd (spacetime)
dimensions. In the second part we study global anomalies at the boundary of BTI states in even dimensions. 
In a previous work [Phys.~Rev.~B 95, 035149 (2017)] we argued that the boundary of the BTI phase exhibits
a global anomaly which is an analogue of the parity anomaly of Dirac fermions in three dimensions. Here we elevate this argument
to a proof for the boundary of the two-dimensional BTI state by explicitly computing the partition function
of the gauged NLSM describing the boundary. We then use the powerful equivariant localization technique to show that this global 
anomaly is robust against all smooth deformations of the target space of the NLSM which preserve the $U(1)\rtimes\mathbb{Z}_2$ 
symmetry of the BTI state. We also comment on the difficulties of generalizing this latter proof to higher dimensions.
Finally, we discuss the expected low energy behavior of the boundary theories studied in this paper when the coupling
constants are allowed to flow under the renormalization group.

\end{abstract}

\pacs{}

\maketitle

\section{Introduction}
\label{sec:intro}

In the past few years it was realized that a powerful way to understand symmetry-protected topological (SPT) phases with 
symmetry group $G$ in $d$ (spacetime) dimensions is to study \emph{`t Hooft anomalies} of $(d-1)$-dimensional theories with
global $G$-symmetry~\cite{wen2013classifying,kapustin2014symmetry,kapustin2014anomalies,kapustin2014anomalous}. A 
theory with global $G$-symmetry has a `t Hooft anomaly if it cannot be consistently coupled to 
a background gauge field $A$ for the symmetry group $G$~\cite{hooft1980naturalness}. 
It is often the case that an anomalous $(d-1)$-dimensional theory 
can be realized in a gauge invariant manner at the boundary of a $d$-dimensional SPT phase. In that case, the anomaly of the 
boundary theory is canceled by the gauge variation of the bulk effective action for the SPT phase. This cancellation mechanism is 
known as \emph{anomaly inflow}~\cite{callan1985anomalies}. It is likely that all bulk-boundary correspondences in SPT phases can 
be understood through some version of the anomaly inflow mechanism, but perhaps involving
global anomalies instead of the perturbative anomalies originally studied in Ref.~\onlinecite{callan1985anomalies}. 

It is clear from the discussion above that characterizing boundary anomalies offers a precise way to understand the 
bulk-boundary correspondence in SPT phases, topological insulators, and related systems. For example, the presence of a single 
chiral fermion at the edge of the $\nu=1$ integer quantum Hall state in $2+1$ dimensions (and also the single chiral boson at
the edge of the Laughlin states) can be understood very simply using anomaly inflow 
arguments~\cite{wen1991gapless,stone1991edge}. 
This chiral fermion is needed to cancel the gauge variation of the bulk Chern-Simons 
term\footnote{We use differential form notation and work in a system of units where $\hbar=e=c=1$.}
\beq
	S_{CS}[A]= \frac{1}{4\pi}\int_{X}A\wedge dA\ ,
\eeq
which describes the response of the integer quantum Hall state to an external electromagnetic field $A=A_{\mu}dx^{\mu}$. 
We also note that anomaly inflow has been discussed for analogs of the integer quantum Hall state in all odd spacetime
dimensions~\cite{karabali2006quantum}.

A related, but much more subtle, example of anomaly inflow occurs in time-reversal invariant, free-fermion 
topological insulators in $3+1$ dimensions~\cite{fu2007topological,QHZ2008}. 
In Ref.~\onlinecite{witten2015fermion} Witten has shown (among other results) that 
the bulk-boundary correspondence in this system can be understood very precisely in terms of the parity 
anomaly of a Dirac fermion with $U(1)$ and time-reversal symmetry in $2+1$ 
dimensions~\cite{Redlich,niemi1983axial,alvarez1985anomalies,witten2016parity}. The parity anomaly is intimately related
to the Atiyah-Potodi-Singer index theorem~\cite{APS1,APS2,APS3} for the Dirac operator on an even-dimensional manifold with
boundary (see Ref.~\onlinecite{alvarez1985anomalies} for the relation), and this connection was a central theme in 
Ref.~\onlinecite{witten2015fermion}. The connection between the parity anomaly and the boundary 
theory of the topological insulator, and in particular the fact that the bulk and boundary together are gauge invariant, was also 
previously discussed in Ref.~\onlinecite{mulligan2013topological}.

In a separate series of developments, bosonic analogues of the integer quantum Hall and topological insulator states were 
introduced and studied in detail in the SPT literature. 
The bosonic integer quantum Hall (BIQH) state is an SPT phase of bosons with $U(1)$ symmetry in $2+1$
dimensions~\cite{SenthilLevin,VL2012,grover2013,regnault2013,he2015,moller2009,moller2015,ye2013,liu2014,furukawa2013integer,wu2013quantum}.
It is characterized by a Hall conductance which is an \emph{even} integer (in units of $\frac{e^2}{h}$). 
On the other hand, the bosonic topological insulator (BTI) state is an SPT phase of bosons with $U(1)$ symmetry and $\mathbb{Z}_2$ time-reversal
symmetry in $3+1$ dimensions~\cite{VS2013,MKF2013,SenthilWang2013,ye2015vortex}. It is characterized by a bulk 
electromagnetic response of the ``Chern character" type
\beq
	S_{CC}[A]= \frac{\Theta}{8\pi^2}\int_{X}F\wedge F\ ,
\eeq
with coefficient $\Theta=2\pi$. In a recent work, the present authors computed the electromagnetic response of generalizations of 
the BIQH and BTI states to \emph{all} odd and even spacetime dimensions, respectively~\cite{lapa2017}.

Given these separate developments, a natural next step would be to give a precise characterization of the anomalies at the 
boundaries of the BIQH and BTI states. In Ref.~\onlinecite{lapa2017} we initiated such a program. There we used a nonlinear
sigma model (NLSM) 
description~\cite{CenkeClass1,CenkeClass2,CenkeBraiding,CenkeLine,CenkeSU2N,CenkeDual,xu2013wave,ElseNayak,liu2013symmetry,
you2016stripe,you2016decorated} of the boundary of the BIQH state in odd dimensions to compute the perturbative $U(1)$ anomaly of the boundary 
theory. 
Our result implied that the electromagnetic response of the bulk of a BIQH
state in $2m-1$ dimensions is characterized by a Chern-Simons term\footnote{See Eq.~\eqref{eq:CS-term-higher-dim} for our
normalization of the Chern-Simons term in $2m-1$ dimensions.} with level $N_{2m-1}= (m!)k$,
$k\in\mathbb{Z}$, where the value $k=1$ represents the fundamental BIQH state.

In Ref.~\onlinecite{lapa2017} we also argued that the boundary theory of the $2m$-dimensional BTI state exhibits 
a bosonic analogue of the well-known parity\footnote{As we explained in Ref.~\onlinecite{lapa2017}, in spacetime dimensions $2m$ with $m$ 
\emph{odd} the $\mathbb{Z}_2$ symmetry of the BTI state is a unitary charge-conjugation symmetry and not time-reversal
symmetry. For these cases the word ``parity"
is not a very good description of the symmetry which is anomalous. However, for ease of presentation we will continue to refer
to the global anomalies discussed here as ``bosonic analogues of the parity anomaly".} anomaly of
Dirac fermions in three dimensions. Our argument was based on a demonstration (again
using a NLSM description) that the boundary of the BTI state can exhibit a $\mathbb{Z}_2$ symmetry-breaking electromagnetic 
response described by a Chern-Simons term with level $N_{2m-1}=\frac{m!}{2}$ for the external field $A$. Since 
this boundary response is \emph{half} the response of 
the fundamental BIQH state in $2m-1$ dimensions, we argued, by analogy with the case of a massless Dirac fermion 
(with Hall conductance = Chern-Simons level = $\frac{1}{2}$) on the surface of the $(3+1)$-dimensional 
topological insulator~\cite{fu2007topological,QHZ2008}, that 
the boundary theory of the BTI displays a bosonic analogue of the parity anomaly.

In this paper we continue this program of characterizing anomalies at the boundary of BIQH and BTI states. In the first
part of the paper we revisit the perturbative $U(1)$ anomaly at the boundary of $(2m-1)$-dimensional BIQH states. In 
Ref.~\onlinecite{lapa2017} we computed this anomaly by gauging the Wess-Zumino (WZ) term in an $O(2m)$ NLSM description of the 
boundary of the BIQH state. In any NLSM, the field is a map from spacetime to a manifold $\mathcal{M}$, known as the 
\emph{target space} of the NLSM. In the $O(2m)$ NLSM the target space is just the $(2m-1)$-dimensional unit sphere 
$S^{2m-1}$, and the NLSM field $\mb{n}$ is a $2m$-component unit vector. 
This particular NLSM description possesses a $SO(2m)$ global symmetry, which is much larger 
than the $U(1)$ symmetry required to protect the BIQH state. One might then wonder if (perhaps) more realistic models of the BIQH boundary can be found
which still possess the correct perturbative $U(1)$ anomaly, but have only the $U(1)$ global symmetry of the BIQH state.
In this paper we show that a large family of such models do indeed exist by proving the following result. 

\begin{framed}
Let $\M$ be \emph{any} $(2m-1)$-dimensional manifold which can be reached from $S^{2m-1}$ by 
smooth deformations which preserve the $U(1)$ symmetry of the BIQH phase (i.e., we have a diffeomorphism $f: \M \to S^{2m-1}$ 
which is \emph{equivariant} with respect to the $U(1)$ symmetry). Then a description of the boundary of the BIQH 
state using a NLSM with target space $\M$ has \emph{the same} perturbative $U(1)$ anomaly as the $O(2m)$ NLSM 
description.  
\end{framed} 

In the second part of the paper we revisit the bosonic analogue of the parity anomaly at the boundary of the BTI states. In the
simplest case of the BTI state in two spacetime dimensions we are able to compute the partition function of the gauged boundary theory 
exactly. The BTI state in two dimensions has the symmetry group $G= U(1)\rtimes\mathbb{Z}_2$, where $\mathbb{Z}_2$ 
represents a unitary charge-conjugation symmetry. Our exact computation of the boundary partition function shows that the 
boundary of the BTI does indeed exhibit a bosonic analogue of the global anomaly of Dirac fermions in $0+1$ dimensions which also have $U(1)$
symmetry and $\mathbb{Z}_2$ charge-conjugation symmetry~\cite{elitzur1986origins}.
We first compute this anomaly within the $O(3)$
NLSM description (with target space $S^2$) of the BTI boundary which we previously used in Ref.~\onlinecite{lapa2017}. 
Based on this calculation, one might again wonder if a more realistic model of the BTI boundary can be found which has the same 
global anomaly, but which possesses only the $G= U(1)\rtimes\mathbb{Z}_2$ symmetry of the BTI state and not the full $SO(3)$ 
symmetry of the $O(3)$ NLSM. We again show that such models do exist by proving the following result. 

\begin{framed}
Let $\M$ be \emph{any} two-dimensional manifold which can be reached from $S^2$ by smooth 
deformations which preserve the full $G= U(1)\rtimes\mathbb{Z}_2$ symmetry of the BTI state (i.e., we have a diffeomorphism 
$f: \M \to S^2$ which is \emph{equivariant} with respect to the action of the group $G$).  Then a description of the boundary of 
the BTI state using a NLSM with target space $\M$ has \emph{the same} global anomaly as the $O(3)$ NLSM 
description.  
\end{framed}

To prove this result we use the powerful \emph{equivariant localization} technique originally developed for the exact computation
of certain phase space path 
integrals~\cite{blau1990path,keski1991topological,niemi1992cohomological,niemi1994exact,szabo2003equivariant}. Whereas for the perturbative anomaly we are able to extend our proof to any spacetime dimension, the calculation for global anomalies becomes challenging in higher dimensions and is not easily extendable. We comment on this difficulty later, and discuss possible alternative approaches.

As in our previous work~\cite{lapa2017}, gauged WZ actions play a central role in the calculations in this paper. 
Gauging WZ actions, and also obstructions to gauging these actions (i.e., anomalies), have been discussed previously in 
Refs.~\onlinecite{witten1983global,manes1985,WittenHolo,HullSpence1,HullSpence2,wu1993cohomological,figueroa1,figueroa2}. 
Since we consider two kinds of anomalies in this paper (perturbative and global), it is important for us to explain at the outset 
how exactly our anomalies are related to obstructions to gauging a WZ action. For the perturbative $U(1)$ anomalies that we 
study, the anomaly that we find is a direct result of the existence of an obstruction to gauging the WZ action. Therefore, 
these anomalies are already present at the level of the classical action for these theories. On the other hand, for the 
global anomalies that we study there is \emph{no obstruction} to gauging the $U(1)$ symmetry of the WZ action. Instead, the 
anomaly is a completely quantum effect which stems from an inability to regulate the theory in such a way as to preserve
both large $U(1)$ gauge invariance, and the additional discrete $\mathbb{Z}_2$ symmetry of the theory.

This paper is organized as follows. In Sec.~\ref{sec:pert} we analyze perturbative $U(1)$ anomalies at the even-dimensional 
boundary of BIQH states in generic odd spacetime dimensions. In Sec.~\ref{sec:global} we analyze the global anomaly at the 
$(0+1)$-dimensional boundary of the $(1+1)$-dimensional BTI state.  In Sec.~\ref{sec:RG-flows} we comment on 
the expected behavior of the boundary theories studied in this paper under renormalization group flows. In
Sec.~\ref{sec:conclusion} we present our conclusions. In Appendix~\ref{app:symp-geom} we review the form of the phase 
space path integral for Hamiltonian systems on a general phase space $\M$ equipped with symplectic form $\omega$. In 
Appendix~\ref{app:EL} we give a brief introduction to the equivariant localization technique for phase space path integrals. Finally, in 
Appendix~\ref{app:dets} we present the detailed calculations of the regularized determinants which appear in the expression 
(obtained from the equivariant localization technique) for the partition function of the BTI boundary. 

\section{Perturbative anomalies in bosonic integer quantum Hall states}
\label{sec:pert}

In this section we study perturbative $U(1)$ anomalies at the boundary of a class of bosonic SPT phases in odd spacetime
dimensions which are protected by the symmetry of the group $G=U(1)$. We refer to these phases as
bosonic integer quantum Hall (BIQH) states. They are all higher-dimensional generalizations of 
the $(2+1)$-dimensional BIQH state introduced in Ref.~\onlinecite{SenthilLevin}. Upon coupling to a background $U(1)$ gauge 
field $A= A_{\mu} dx^{\mu}$, the boundary of these states exhibits a perturbative $U(1)$ anomaly. For the BIQH phase
in $2m-1$ dimensions, the anomaly of the boundary is such that it can be compensated by a bulk Chern-Simons (CS) term 
\beq
	S_{CS}[A]= \frac{N_{2m-1}}{(2\pi)^{m-1}m!}\int_{X}A\wedge (dA)^{m-1}\  \label{eq:CS-term-higher-dim}
\eeq
with the level $N_{2m-1}$ of the CS term quantized in integer multiples of $m!$ (factorial).
Here $X$ denotes the $(2m-1)$-dimensional bulk spacetime. We computed this anomaly in Ref.~\onlinecite{lapa2017}
using a NLSM description of the boundary theory of the BIQH state. Specifically, we modeled the boundary using
an $O(2m)$ NLSM with Wess-Zumino (WZ) term, with a particular action of the group $U(1)$ on the NLSM field. 
The field in this model is a $2m$-component unit vector $\mb{n}= (n^1,\dots,n^{2m})$, and so the target space of 
the $O(2m)$ NLSM is the $(2m-1)$-dimensional unit sphere $S^{2m-1}$. 

In this section we first recall the result of Ref.~\onlinecite{lapa2017}, and we also show that the anomaly
computed there is well-defined in the sense that it is independent of a certain freedom in the specific form of the terms appearing 
in the gauged WZ action for the boundary theory. We then consider alternative descriptions of the BIQH state using NLSMs with
a general target space $\mathcal{M}$, and we prove that if $\M$ can be obtained from $S^{2m-1}$ by smooth deformations
which preserve the $U(1)$ symmetry of the BIQH state, then the anomaly of the NLSM theory with target space $\mathcal{M}$ 
is \emph{identical} to the anomaly of the $O(2m)$ NLSM theory.  Later in the paper, in Sec.~\ref{sec:RG-flows}, 
we discuss the expected low energy behavior of the NLSMs discussed in this section.

The results of this section prove that the anomaly computed in Ref.~\onlinecite{lapa2017} is robust against arbitrary smooth, 
symmetry-preserving deformations of the NLSM used to describe the boundary of the BIQH state. This is exactly what one 
hopes for in a model of an SPT phase: smooth, symmetry-preserving deformations of a model of an SPT phase should not affect
the ability of that model to capture the universal properties of the SPT phase, provided that the deformations do not take one across
a phase boundary.
We also note here that in Ref.~\onlinecite{lapa2017} we gave a more general \emph{gauge invariance} argument for the 
quantization of the level $N_{2m-1}$ of the CS term describing the bulk response of the BIQH state. That
argument also implies that the boundary anomaly is robust and independent of the specific details of any particular model
of the boundary of the BIQH state. Therefore, the results of this section could have been anticipated from the gauge 
invariance argument in Ref.~\onlinecite{lapa2017}. However, it is also instructive to have an explicit proof of this invariance
for the class of NLSM descriptions of the boundary considered here. 

\subsection{Review of $O(2m)$ NLSM calculation of the anomaly}

We start by reviewing the calculation of the boundary anomaly of the BIQH state using the $O(2m)$ NLSM description. 
The boundary of the $(2m-1)$-dimensional BIQH state can be described by an $O(2m)$ NLSM with WZ term. Let $X_{bdy}$
denote the $(2m-2)$-dimensional boundary spacetime. As we discussed above, the NLSM field 
$\mb{n}= (n^1,\dots,n^{2m})$ should be understood as a map 
$\mb{n}: X_{bdy}\to S^{2m-1}$ from the boundary spacetime $X_{bdy}$ to the target space of the
NLSM, which is just the unit sphere $S^{2m-1}$ in this case. 

The WZ term for the NLSM requires the following ingredients for its 
construction. First, we need the volume form $\omega_{2m-1}$ on $S^{2m-1}$. In terms of the coordinates 
$n^a$, $a=1,\dots,2m$, it takes the form
\beq
	\omega_{2m-1}= \sum_{a=1}^{2m}(-1)^{a-1} n^a dn^1\wedge \cdots \wedge \overline{dn^a}\wedge \cdots dn^{2m}\ ,
\eeq
where the overline means to omit that term from the wedge product.
Next, we need an extension $\B$ of the boundary spacetime $X_{bdy}$ such that $\pd \B= X_{bdy}$, where $\pd\B$ denotes
the boundary of $\B$. Finally, we need an extension $\tilde{\mb{n}}$ of the NLSM field $\mb{n}$ into the bulk of $\B$ such
that $\tilde{\mb{n}}|_{\pd\B}= \mb{n}$. The extended field $\tilde{\mb{n}}$ should be viewed as a map
$\tilde{\mb{n}}: \B \to S^{2m-1}$. Then the WZ term for the $O(2m)$ NLSM on the $(2m-2)$-dimensional boundary spacetime
$X_{bdy}$ takes the form
\beq
	S_{WZ}[\mb{n}]= \frac{2\pi k}{\A_{2m-1}}\int_{\mathcal{B}}\tilde{\mb{n}}^*\omega_{2m-1}\ ,
\eeq
where $k\in\mathbb{Z}$ is the level of the WZ term and $\A_{2m-1}=\text{Area}[S^{2m-1}] = \frac{2\pi^m}{(m-1)!}$. 
Here the notation $\tilde{\mb{n}}^*\omega_{2m-1}$ denotes the pullback of the volume form $\omega_{2m-1}$ on 
$S^{2m-1}$ to the extended boundary spacetime $\B$ via the map $\tilde{\mb{n}}: \B \to S^{2m-1}$. 

The WZ term can be written in a more familiar form if we introduce a system of local coordinates $(s,x^0,\dots,x^{2m-3})$ on $\B$,
where $(x^0,\dots,x^{2m-3})$ are a system of local coordinates on $X_{bdy}$, and where $s\in[0,1]$ is a coordinate
for the extra direction in $\B$. We choose boundary conditions on the extended field configuration 
such that $\tilde{\mb{n}}$  is equal to a trivial constant configuration at $s=0,$ and $\tilde{\mb{n}}=\mb{n}$ at $s=1.$ Hence,
the physical boundary spacetime $X_{bdy}$ is located at $s=1$. In these coordinates the WZ term takes the
more explicit form
\begin{widetext}
\beq
	S_{WZ}[\mb{n}]= \frac{2\pi k}{\mathcal{A}_{2m-1}}\int_0^1 ds \int d^{2m-2}x\ \ep_{a_1\cdots a_{2m}} \tilde{n}^{a_1}\pd_{s}\tilde{n}^{a_2}\pd_{x^0}\tilde{n}^{a_3}\cdots\pd_{x^{2m-3}}\tilde{n}^{a_{2m}}\ ,
\eeq
\end{widetext}
where we sum over all indices which appear once as a subscript and once as a superscript (the standard summation notation).
In addition to the WZ term, the action for the $O(2m)$ NLSM also includes a conventional kinetic term
\beq
	S_{kin}[\mb{n}]= \frac{1}{2f}\int d^{2m-2}x\ (\pd_{\mu}\mb{n})\cdot(\pd^{\mu}\mb{n})\ , \label{eq:kinetic-term-O(N)}
\eeq
where $f$ is a coupling constant with dimensions of $(mass)^{4-2m}$ (the power is equal to two minus the boundary
spacetime dimension).

The action of the $U(1)$ symmetry that protects the BIQH state on the NLSM field is best described by first pairing the components of $\mb{n}$ 
into $m$ ``bosons" $b_{\ell}= n^{2\ell-1} + in^{2\ell}$, $\ell=1,\dots,m$. Then, for the NLSM model of the BIQH phase, the 
$U(1)$ symmetry can be defined to act on these bosons as~\cite{SenthilLevin,lapa2017}
\beq
	U(1): b_{\ell} \to e^{i\xi}b_{\ell}\ ,\ \forall\ \ell\ . \label{eq:U1-sym}
\eeq
Let us briefly explain the rationale for this choice of the $U(1)$ action. In the NLSM description of bosonic SPT phases from
Ref.~\onlinecite{CenkeClass1}, the 
information about the symmetry group $G$ is encoded in a homomorphism $\sigma: G \to SO(2m)$ (in the case of unitary 
symmetries which have trivial action on spacetime). The NLSM equipped with the homomorphism $\sigma$ will describe a
trivial phase if there exists a vector $\mb{v}$ such that $\sigma(g)\mb{v}=\mb{v}$, $\forall\ g\in G$. This is because in this
case it is possible to add a ``Zeeman" term $\mb{n}\cdot\mb{v}$ to the NLSM action to drive the NLSM into a trivial direct 
product state in which $\mb{n}$ is parallel or anti-parallel to $\mb{v}$ at all points in space. Therefore, we must choose
a homomorphism $\sigma$ where no such vector $\mb{v}$ exists if we want our NLSM to describe a nontrivial SPT phase with $U(1)$ 
symmetry. Mathematically, the problem is to embed $U(1) \cong SO(2)$ inside the maximal torus of $SO(2m)$ in such a way that
no vector $\mb{v}$ is fixed under the action of $\sigma(g)$ $\forall\ g\in U(1)$. 
The unique solution to this problem\footnote{More precisely, this is the unique solution if we demand that the fundamental 
particles in the model carry unit electric charge.}, modulo trivial permutations of the components $n^a$ in the definition of the bosons 
$b_{\ell}$, is the one in Eq.~\eqref{eq:U1-sym}. 

Next, we couple the NLSM describing the boundary of the BIQH state to a background $U(1)$ gauge field 
$A= A_{\mu} dx^{\mu}$, and attempt to construct an action which is invariant under the gauge transformation 
\beqa
	b_{\ell} &\to& e^{i\xi}b_{\ell}\ ,\ \forall\ \ell\  \nnb \\
	A &\to& A + d\xi\ ,
\eeqa
where $\xi$ is now a function of the boundary spacetime coordinates. 
This gauge transformation can be recast in a more geometric form using the
vector field $\un{v}= v^a\frac{\pd}{\pd n^a}$ which generates the action of the $U(1)$ symmetry on $S^{2m-1}$. 
Concretely, this means that under an infinitesimal $U(1)$ transformation, the coordinates on $S^{2m-1}$ transform as
\beq
	n^a \to n^a + \xi v^a\ .
\eeq
For the $U(1)$ symmetry action defined in Eq.~\eqref{eq:U1-sym}, the vector field $\un{v}$ takes the form
\beq
	\underline{v}= \sum_{\ell=1}^m \left( -n^{2\ell}\frac{\pd}{\pd n^{2\ell-1}} + n^{2\ell-1}\frac{\pd}{\pd n^{2\ell}} \right)\ .
\eeq
This transformation of the coordinates also induces a transformation for general $p$-forms $\beta$ on $S^{2m-1}$,
\beq
	\beta \to \beta + \mathcal{L}_{\xi\un{v}}\beta\ ,
\eeq
where $\mathcal{L}_{\un{v}}= di_{\un{v}} + i_{\un{v}}d$ is the Lie derivative (acting on differential forms) along $\un{v}$,
and $i_{\un{v}}$ is the interior multiplication by $\un{v}$ ($d$ is the ordinary exterior derivative).

To simplify the presentation of the gauged WZ action it is best to work with a more compact notation. Let us define the 
normalized volume form $\al^{(2m-1)}= \frac{\omega_{2m-1}}{\A_{2m-1}}$ so that the WZ term can be written as
\beq
	S_{WZ}[\mb{n}] = 2\pi k\int_{\mathcal{B}}\tilde{\mb{n}}^*\al^{(2m-1)}\ .
\eeq
The derivation of the gauged WZ action is somewhat technical, and so we refer the reader to 
Ref.~\cite{lapa2017} for details.
In Ref.~\cite{lapa2017} we showed that the gauged WZ action for the $O(2m)$ NLSM takes the form 
\begin{align}
	S_{WZ,gauged}[\mb{n},&A] =  \nnb \\
S_{WZ}[\mb{n}]+ 2\pi k &\sum_{r=1}^{m-1}\int_{X_{bdy}}A\wedge F^{r-1}\wedge\mb{n}^*\al^{(2m-1-2r)}\ , \label{eq:gauged-WZ-action}
\end{align}
where the $\al^{(2m-1-2r)}$ are a set of differential forms on $S^{2m-1}$ of degree $2m-1-2r$, $r=1,\dots,m-1$, which have 
a form that we now discuss. 

First, for each $\ell=1,\dots, m$, we define one-forms $\mathcal{J}_{\ell}$ and two-forms $\mathcal{K}_{\ell}$ on $S^{2m-1}$ 
by
\begin{subequations}
\label{eq:J-K-def}
\beqa
	\mathcal{J}_{\ell} &=& n_{2\ell-1}dn_{2\ell}- n_{2\ell}dn_{2\ell-1}  \\
	\mathcal{K}_{\ell} &=& dn_{2\ell-1}\wedge dn_{2\ell}\ .
\eeqa
\end{subequations}
Then, for each $r= 0,\dots,m-1,$  we define the forms $\Omega^{(r)}$ by
\beq
	\Omega^{(r)}= \sum_{\ell_1,\dots,\ell_{m-r}=1}^m \mathcal{J}_{\ell_1}\wedge \mathcal{K}_{\ell_2}\wedge \dots \wedge \mathcal{K}_{\ell_{m-r}}\ . \label{eq:omega-forms}
\eeq
In particular, $\Omega^{(r)}$ is a form of degree $2m-1-2r$ and the volume form can be expressed in terms of $\Omega^{(0)}$ 
as $\omega_{2m-1}= \frac{1}{(m-1)!}\Omega^{(0)}$. One can show that these forms obey the relation
\beq
	i_{\underline{v}}\Omega^{(r)} = \frac{1}{2}d\Omega^{(r+1)}\ , \label{eq:int-mult-BIQH}
\eeq
and this relation allows for the construction of the gauged WZ action. 
In terms of these forms, the forms $\al^{(2m-1-2r)}$ appearing in the gauged WZ action are given by 
\beq
	\al^{(2m-1-2r)}= \frac{1}{\A_{2m-1}}\frac{1}{(m-1)!}\frac{1}{2^r}\Omega^{(r)}\ . \label{eq:alphas}
\eeq
This collection of forms obeys the set of equations
\beq
	i_{\un{v}}\al^{(2m-1-2r)} = d\al^{(2m-1-2r-2)}\ ,\ r=0,\dots,m-2 \ ,
\eeq
and 
\beq
	i_{\un{v}}\al^{(1)}= \frac{1}{\A_{2m-1}}\frac{1}{(m-1)!}\frac{1}{2^{m-1}}\ .
\eeq
Since $\A_{2m-1}= \frac{2\pi^m}{(m-1)!}$, we can rewrite the equations satisfied by the $\al^{2m-1-2r}$ as
\bseq
\label{eq:descent-eqns}
\begin{align}
	i_{\un{v}}\al^{(2m-1-2r)} &=  d\al^{(2m-1-2r-2)}\ ,\ r=0,\dots,m-2\ , \label{eq:descent-eqns-higher-forms} \\ 
	i_{\un{v}}\al^{(1)} &= \frac{1}{(2\pi)^m}\ .
\end{align}
\eseq

Under a $U(1)$ gauge transformation $b_{\ell}\to e^{i\xi}b_{\ell}$, $A\to A+d\xi$, the gauged WZ action for the 
$O(2m)$ NLSM transforms as
\beq
	\delta_{\xi} S_{WZ,gauged}[\mb{n},A]= k \int_{X_{bdy}} \xi \left(\frac{F}{2\pi}\right)^{m-1}\ . \label{eq:BIQH-anomaly}
\eeq
In the $O(2m)$ NLSM description of the boundary of the BIQH state, this anomaly of the gauged WZ term 
implies that the topological electromagnetic response of the bulk of the BIQH 
state is described by a CS term with level $N_{2m-1}= -(m!)k$, i.e., the level must be an integer multiple of $m!$. By inspecting the individual terms in the gauged WZ action, one can see that the anomaly in Eq.~\eqref{eq:BIQH-anomaly} 
is completely determined by the value of $i_{\un{v}}\al^{(1)}$ as shown in Eqs.~\eqref{eq:descent-eqns}. This is because 
Eq.~\eqref{eq:descent-eqns-higher-forms} guarantees that the transformation of the form 
$\al^{(2m-1-2r)}$ in the $r^{th}$ term in Eq.~\eqref{eq:gauged-WZ-action} is canceled by the transformation of the gauge 
field $A$ in the $(r+1)^{th}$ term. This means that the final anomaly only depends on the transformation of $\al^{(1)}$ in the 
$(m-1)^{th}$ term (i.e., the last term). 
It turns out that the equations which define the form $\al^{(1)}$ do not have a unique solution, and 
in the computation above we have chosen a particular solution. We now show that although there is an ambiguity in the
choice of solution for $\al^{(1)}$, the anomaly of the gauged action is not affected by this ambiguity.

\subsection{Uniqueness of the anomaly}

In the previous subsection we showed that the anomaly of the $O(2m)$ NLSM with WZ term is completely determined by the
one-form $\al^{(1)}$ which appears in the final term of the gauged WZ action, and we also mentioned that 
$\al^{(1)}$ is not unique. If we are to ascribe any physical meaning to the anomaly computed in the last subsection, then we need 
to make sure that the anomaly is not affected by the ambiguity in the choice of the form $\al^{(1)}$. In this section 
we prove that the anomaly is well-defined even though the choice of $\al^{(1)}$ is not unique.

We start by precisely characterizing the ambiguity in the choice of the one-form $\al^{(1)}$. According to 
Eqs.~\eqref{eq:descent-eqns}, this form should satisfy the equation
\beq
	i_{\un{v}}\al^{(3)}= d\al^{(1)}\ . \label{eq:al1-eqn}
\eeq
However, for a given three-form $\al^{(3)}$, the solutions to this equation for $\al^{(1)}$ are not unique. To see this, 
let us fix a choice of $\al^{(3)}$ (and also $\al^{(5)},\dots,\al^{(2m-3)}$) and suppose that we have two solutions 
$\al^{(1)}$ and $\tilde{\al}^{(1)}$ to 
Eq.~\eqref{eq:al1-eqn}. If we subtract the equation for $\al^{(1)}$ from the equation for $\tilde{\al}^{(1)}$ then we find that 
these two forms are related by the equation
\beq
	d(\tilde{\al}^{(1)}-\al^{(1)})= 0\ ,
\eeq
i.e., the difference $\tilde{\al}^{(1)}-\al^{(1)}$ is a closed form on $S^{2m-1}$. However, on the sphere $S^{2m-1}$ all
closed one-forms are also exact\footnote{On $S^{2m-1}$ the de Rham cohomology groups $H^r_{dR}(S^{2m-1})$ are trivial for 
$r=1,\dots,2m-2$.}, which means that we have
\beq
	\tilde{\al}^{(1)}-\al^{(1)}= d\gamma^{(0)}
\eeq
for some function $\gamma^{(0)}$ on $S^{2m-1}$. 

We now want to understand the possible dependence of the anomaly on the function $\gamma^{(0)}$ which parametrizes
the ambiguity in the solution for $\al^{(1)}$. Therefore we should compare the gauged WZ action 
constructed using $\al^{(1)}$ with the gauged WZ action constructed using $\tilde{\al}^{(1)}$ (but keeping all other terms
in the action the same). Let $S_{WZ,gauged}[\mb{n},A]$ be the gauged WZ action constructed
using the form $\al^{(1)}$, and let $\tilde{S}_{WZ,gauged}[\mb{n},A]$ be the gauged WZ action constructed from the
form $\tilde{\al}^{(1)}$. These actions differ by a single term
\begin{align}
	\tilde{S}_{WZ,gauged}[\mb{n},A] -  &S_{WZ,gauged}[\mb{n},A] \nnb \\
	 &= 2\pi k\int_{X_{bdy}} A\wedge F^{m-2} \wedge \mb{n}^*d\gamma^{(0)} \nnb \\
	&= 2\pi k\int_{X_{bdy}}\mb{n}^*\gamma^{(0)}\ F^{m-1}\ ,
\end{align}
where we rearranged the forms and performed an integration by parts to derive the second equality. Under a gauge transformation this
difference transforms as
\begin{align}
	\delta_{\xi}\tilde{S}_{WZ,gauged}[\mb{n},A] - \delta_{\xi}S_{WZ,gauged}[&\mb{n},A] \nnb \\
= 2\pi k\int_{X_{bdy}}&\mb{n}^*(\mathcal{L}_{\xi\un{v}}\gamma^{(0)})\ F^{m-1}\ .
\end{align}
However, since $\gamma^{(0)}$ is a \emph{function}, we have
\beqa
	\mathcal{L}_{\xi\un{v}}\gamma^{(0)} &=& d(\xi i_{\un{v}} \gamma^{(0)}) + \xi i_{\un{v}}d\gamma^{(0)} \nnb \\
	&=& \xi i_{\un{v}}d\gamma^{(0)} \nnb \\
	&=& \xi \mathcal{L}_{\un{v}}\gamma^{(0)}\ ,
\eeqa
where we used the fact that $i_{\un{v}}\gamma^{(0)}=0$. Then
the difference of gauge transformations reduces to 
\begin{align}
	\delta_{\xi}\tilde{S}_{WZ,gauged}[\mb{n},A] - \delta_{\xi}S_{WZ,gauged}[&\mb{n},A] \nnb \\
= 2\pi k\int_{X_{bdy}}&\xi\  \mb{n}^*(\mathcal{L}_{\un{v}}\gamma^{(0)})\ F^{m-1}\ .
\end{align}

We can now make the following observation. The gauged action $S_{WZ,gauged}[\mb{n},A]$ constructed using 
$\al^{(1)}$ from Eq.~\eqref{eq:alphas} still possesses \emph{global} $U(1)$ symmetry and, in particular, is invariant under the 
transformation $b_{\ell}\to e^{i\xi} b_{\ell}$ for an infinitesimal \emph{constant} parameter $\xi$. However, the above 
considerations show that under the same infinitesimal $U(1)$ transformation, the gauged action 
$\tilde{S}_{WZ,gauged}[\mb{n},A]$ constructed from $\tilde{\al}^{(1)}$ will transform as
\beq
	\delta_{\xi}\tilde{S}_{WZ,gauged}[\mb{n},A]= 2\pi k\xi\int_{X_{bdy}}  \mb{n}^*(\mathcal{L}_{\un{v}}\gamma^{(0)})\ F^{m-1}\ .
\eeq
Now even if the gauged WZ action cannot be made to be invariant under $U(1)$ gauge transformations, we should still require
it to be invariant under \emph{global} $U(1)$ transformations. Therefore we must demand that for any alternative
solution $\tilde{\al}^{(1)}$ to Eq.~\eqref{eq:al1-eqn}, the function $\gamma^{(0)}$ relating this form to $\al^{(1)}$ 
from Eq.~\eqref{eq:alphas} should satisfy
\beq
	\mathcal{L}_{\un{v}}\gamma^{(0)}= 0\ ,
\eeq
i.e., this function should be invariant under the action of the $U(1)$ symmetry on $S^{2m-1}$. Then, since we have the
relation $\mathcal{L}_{\xi\un{v}}\gamma^{(0)}= \xi \mathcal{L}_{\un{v}}\gamma^{(0)}$ for any function 
$\gamma^{(0)}$ and any spacetime-dependent $\xi$, we immediately find that the anomaly of the gauged WZ action is
not sensitive to the ambiguity in the choice of $\al^{(1)}$. In other words, the requirement that the gauged WZ action 
should still possess \emph{global} $U(1)$ symmetry is enough to ensure that the anomaly of the gauged action is well-defined
and independent of the ambiguity in the choice of $\al^{(1)}$.

\subsection{Deforming the target space}

Now that we know that the anomaly in Eq.~\eqref{eq:BIQH-anomaly} is well-defined, we can move on and study how
deformations of the target space of the NLSM might affect the anomaly. Recall that we previously derived this
anomaly using the $O(2m)$ NLSM with target space $S^{2m-1}$. In this subsection we show that this anomaly is not affected by
arbitrary smooth, symmetry-preserving deformations of the target space of the 
NLSM. The notion of a smooth, symmetry-preserving deformation of the target space can be formulated precisely in terms of
diffeomorphisms which are \emph{equivariant} with respect to the symmetry action, as we discuss below.

In the NLSM description of the BIQH state the target space $S^{2m-1}$ of the $O(2m)$ NLSM is equipped with an 
action of the group $U(1)$. For any $g\in U(1)$ let us write $g\cdot \mb{n}$ to denote the image of the point $\mb{n} \in S^{2m-1}$ 
under the action of the group element $g$. As we discussed above, the $U(1)$ action on $S^{2m-1}$ is generated by the vector
field $\un{v}$ in the sense that $n^a \to n^a + \xi v^a$ under an infinitesimal $U(1)$ transformation parametrized by 
$\xi$.  Now suppose that $\M$ is another $(2m-1)$-dimensional manifold with the following properties.
\begin{enumerate}[{(1)}]
\item There is a $U(1)$ action on $\M$ generated by a vector field $\un{w}$. \label{prop1}
\item There exists a Riemannian metric on $\M$ for which $\un{w}$ is a Killing vector.
\item There exists a diffeomorphism $f: \M \to S^{2m-1}$ which is \emph{equivariant} with respect to the $U(1)$ 
action, i.e.,
\beq
	g\cdot f(\mb{m})= f(g\cdot \mb{m})\ ,\ \forall\ \mb{m}\in \M\ ,\ \forall\ g\in U(1)\ . \label{eq:g-action}
\eeq 
\end{enumerate}
Intuitively, these properties imply that the manifold $\M$ also has a $U(1)$ symmetry, and that it can be reached from $S^{2m-1}$ 
(or vice-versa) by smooth deformations which respect the $U(1)$ symmetry. 
We now show that for any such manifold $\M$ the NLSM with target space $\M$, WZ term at level $k$, and
$U(1)$ action generated by $\un{w}$ possesses the exact same perturbative $U(1)$ anomaly as 
the $O(2m)$ NLSM with WZ term at level $k$.

Before presenting the proof, we first discuss some consequences of the three properties of the map
$f$. First, properties (1) and (2) together imply that we can construct a WZ term for the NLSM with target space $\M$ with
the property that the WZ term is invariant under the $U(1)$ transformation generated by $\un{w}$ (we construct the
WZ term using the volume form on $\M$ determined by its $U(1)$-symmetric Riemannian metric). 
Next, the first part of property (3), namely the fact that $f:\M\to S^{2m-1}$ is a diffeomorphism, 
implies that the de Rham cohomology groups of $\M$ and $S^{2m-1}$ are identical. In addition, the fact that $f$ is a 
diffeomorphism implies that the \emph{degree} of $f$, defined via the equation
\beqa
	\frac{1}{\A_{2m-1}}\int_{\M} f^*\omega_{2m-1} &=& \text{deg}[f]\frac{1}{\A_{2m-1}}\int_{S^{2m-1}} \omega_{2m-1} \nnb \\
	&=& \text{deg}[f]\ ,
\eeqa
is equal to plus or minus one, $\text{deg}[f]=\pm 1$ (see Ch.~VI of Ref.~\onlinecite{flanders1963differential} for the
definition of the degree of a smooth map). Intuitively this means that the map $f$ ``wraps" $\M$ around 
$S^{2m-1}$ only once. This has to be the case since $f$ is injective ($f$ is invertible so it is both injective 
and surjective). In what follows we assume $\text{deg}[f]=1$ so that $f$ is orientation-preserving. This then 
implies that 
\beq
	f^*\left(\frac{\omega_{2m-1}}{\A_{2m-1}}\right)= \frac{\omega_{\M}}{\A_{\M}}\ , \label{eq:volume-form-map}
\eeq
where $\omega_{\M}$ is the volume form on $\M$ determined by its Riemannian metric, and $\A_{\M}= \int_{\M} \omega_{\M}$ is
the area of $\M$.

Next, properties (1) and (3) together imply that 
\beq
	\un{v}= f_*\un{w}\ ,
\eeq
i.e., the vector field $\un{v}$ which generates the $U(1)$ action on $S^{2m-1}$ is equal to the pushforward, via the map $f$, of 
the vector field $\un{w}$ that generates the $U(1)$ action on $\M$. This can be verified by expanding out both sides of 
Eq.~\eqref{eq:g-action} for an element $g\in U(1)$ which is close to the identity. This property implies the following relation, which 
is central to the proof in this section. If $\al$ is a differential form on $S^{2m-1}$, then we have
\beq
	i_{\un{w}}(f^*\al) = f^*(i_{\un{v}}\al)\ . \label{eq:pull-push}
\eeq
This relation implies that the action of interior multiplication commutes with the action of taking the pullback, provided that
we use $i_{\un{w}}$ when acting on forms on $\M$ and $i_{\un{v}}$ when acting on forms on $S^{2m-1}$.
Again, this relation holds because under our assumptions the vector field $\un{v}$ is equal to the pushforward of $\un{w}$ by the 
map $f$.

Now let us consider an alternative description of the boundary of a BIQH state in terms of a NLSM with target space
$\M$, where $\M$ satisfies the three properties stated above. The field in this NLSM theory, which we denote by $\mb{m}$, is a 
map from the boundary spacetime to the manifold $\M$, $\mb{m}: X_{bdy} \to \M$. We also assume that the transformation of the
NLSM field $\mb{m}$ under the $U(1)$ symmetry of the BIQH state is determined by the $U(1)$ action on $\M$ generated by
the vector field $\un{w}$. For example, under an infinitesimal $U(1)$ transformation parametrized by $\xi$ we have
$m^a \to m^a + \xi w^a$, $\forall\ a$. 
The WZ term for this NLSM is constructed in the same way as for the NLSM with target space $S^{2m-1}$. 
We start with a volume form $\omega_{\M}$ on $\M$ which we assume is obtained from a $U(1)$-symmetric Riemannian
metric on $\M$\footnote{Although we did not discuss it explicitly, our earlier construction of the WZ term for the 
$O(2m)$ 
NLSM also required a $U(1)$-symmetric Riemannian metric for $S^{2m-1}$. In particular, the volume form $\omega_{2m-1}$
is the volume form on $S^{2m-1}$ which is obtained from the natural Riemannian metric on $S^{2m-1}$ induced by
the embedding of $S^{2m-1}$ in $\mathbb{R}^{2m}$. The $U(1)$ symmetry of this metric then implied that the $O(2m)$
NLSM with WZ term possessed a global $U(1)$ symmetry.}
 (which exists by our assumption (2) above). We denote the normalized volume form on $\M$ by 
$\beta^{(2m-1)}= \frac{\omega_{\M}}{\A_{\M}}$, where $\A_{\M}= \int_{\M}\omega_{\M}$.
We also need an extension $\tilde{\mb{m}}$ of the NLSM field $\mb{m}$ into the extended boundary spacetime 
$\mathcal{B}$ such that $\tilde{\mb{m}}|_{\pd\B}=\mb{m}$.  In terms of these quantities,
the WZ term for the NLSM with target space $\M$ can be written in the compact form
\beq
	S_{WZ}[\mb{m}]= 2\pi k\int_{\B}\tilde{\mb{m}}^*\beta^{(2m-1)}\ .
\eeq

We can now attempt to couple $S_{WZ}[\mb{m}]$ to the gauge field $A$ and study the perturbative anomaly of the 
gauged action. We find that the gauged WZ term for the NLSM with target space $\M$ takes the form
\begin{align}
	S_{WZ,gauged}[\mb{m},&A]=  \nnb \\ 
S_{WZ}[\mb{m}] + 2\pi k &\sum_{r=1}^{m-1}\int_{X_{bdy}}A\wedge F^{r-1}\wedge\mb{m}^*\beta^{(2m-1-2r)}\ ,
\end{align}
where the forms $\beta^{(2m-1-2r)}$ on $\M$ are obtained by pulling back the forms $\al^{(2m-1-2r)}$ on $S^{2m-1}$ which
appear in the gauged WZ action for the $O(2m)$ NLSM, 
\beq
	\beta^{(2m-1-2r)}= f^*\al^{(2m-1-2r)}\ .
\eeq
The explicit form of $\al^{(2m-1-2r)}$ was given above in Eq.~\eqref{eq:alphas}. 
Using Eq.~\eqref{eq:pull-push} and the fact that the pullback operation commutes with the exterior derivative, we find
that the forms $\beta^{(2m-1-2r)}$ for $r=0,1,\dots,m-1$, obey the set of equations
\bseq
\label{eq:descent-eqns-on-M}
\begin{align}
	i_{\un{w}}\beta^{(2m-1-2r)} &=  d\beta^{(2m-1-2r-2)}\ ,\ r=0,\dots,m-2\ , \\
	i_{\un{w}}\beta^{(1)} &= \frac{1}{(2\pi)^m}\ .
\end{align}
\eseq
These equations are identical to Eqs.~\eqref{eq:descent-eqns} but with $\un{v}$ replaced by $\un{w}$ and
$\al^{(2m-1-2r)}$ replaced by $\beta^{(2m-1-2r)}$. The form of these equations implies that the NLSM theory with target space 
$\M$ has the exact same perturbative $U(1)$ anomaly as the $O(2m)$ NLSM with target space $S^{2m-1}$. In addition,
our argument for the uniqueness of the anomaly from the previous subsection also applies to the theory with target space $\M$.
This follows from the fact that the de Rham cohomology groups of $\M$ are identical to those of $S^{2m-1}$ as a consequence of 
our assumption (3). Therefore we have shown that the perturbative $U(1)$ anomaly at the boundary of the BIQH state is robust 
against arbitrary smooth, symmetry-preserving deformations of the target space of the NLSM used to describe the BIQH state.

\section{Global anomalies in Bosonic Topological Insulator states}
\label{sec:global}

In this section we study global anomalies at the boundary of a class of bosonic SPT phases which exist in even spacetime
dimensions and are protected by the symmetry of the group $G= U(1)\rtimes\mathbb{Z}_2$. We refer to these phases
as bosonic topological insulator (BTI) phases. They are generalizations to all even-dimensional spacetimes of the BTI phase 
introduced in Ref.~\onlinecite{VS2013}. Note also that the system of bosons studied in Ref.~\onlinecite{berg2011quantized}
can be considered to be an example of a $(1+1)$-dimensional BTI state according to our definition.
In all cases the $U(1)$ symmetry represents a physical charge conservation symmetry, however,
the character of the $\mathbb{Z}_2$ symmetry depends on the specific dimension of spacetime. Let the bulk spacetime dimension
be $2m$ for a positive integer $m$. Then for $m$ odd the $\mathbb{Z}_2$ symmetry is a unitary charge-conjugation symmetry, 
while for $m$ even the $\mathbb{Z}_2$ symmetry is an anti-unitary time-reversal symmetry. 

In Ref.~\onlinecite{lapa2017} we 
argued that the boundary theory of the $2m$-dimensional BTI state exhibits a bosonic analogue of the parity anomaly of a 
Dirac fermion in odd dimensions. Our argument was based on the form of the gauged WZ action in an $O(2m+1)$ NLSM
description of the boundary of these phases. Specifically, we showed that if the NLSM field on the boundary of the BTI condensed
in such a way as to break the $\mathbb{Z}_2$ symmetry but preserve the $U(1)$ symmetry of the BTI phase, then the 
boundary would exhibit a BIQH response with \emph{half-quantized} CS coefficient $N_{2m-1}= \frac{m!}{2}$. We
then argued by analogy with the free fermion topological insulator~\cite{fu2007topological,QHZ2008} that this half-quantized BIQH 
response indicated that the boundary of the BTI phase displays a bosonic analogue of the parity anomaly.

In this section we make this reasoning precise in the special case of the BTI state in $1+1$ spacetime dimensions. In this
case we are able to compute the boundary partition function exactly, and the global anomaly can be seen clearly from our
exact result. We start by reviewing the form of the $O(3)$ NLSM action which 
describes the $(0+1)$-dimensional boundary of this BTI state, including the form of the gauged WZ action which
describes the boundary theory coupled to the external gauge field $A$~\cite{lapa2017}. 
We then explicitly compute the boundary partition 
function and show that it cannot retain both the $\mathbb{Z}_2$ symmetry of the BTI \emph{and} large $U(1)$ gauge invariance, 
i.e., the boundary theory possesses a \emph{global anomaly} in the $\mathbb{Z}_2$ symmetry of the BTI state. We then
consider arbitrary smooth, symmetry-preserving deformations of the target space of the NLSM used to describe the BTI, and
we use the powerful \emph{equivariant localization} (EL) technique to show that the boundary partition function and the global
anomaly are robust against such deformations of the model. We also note here that the global anomaly computed
in this section is very similar to the global anomaly computed in Ref.~\onlinecite{elitzur1986origins} for a single Dirac fermion in 
$(0+1)$-dimensions with $U(1)$ symmetry and unitary $\mathbb{Z}_2$ charge-conjugation symmetry.

\subsection{The BTI state in $1+1$ dimensions and its O(3) NLSM description}

The BTI state in $1+1$ dimensions is an SPT phase of bosons with symmetry group $G= U(1)\rtimes\mathbb{Z}_2$, 
where $U(1)$ represents charge conservation and $\mathbb{Z}_2$ is a unitary charge-conjugation (or particle-hole) symmetry.
The semi-direct product ``$\rtimes$" indicates that the $U(1)$ and $\mathbb{Z}_2$ symmetries do not commute with each
other. The physical signature of the BTI state is that a fractional charge of $\pm \frac{1}{2}$ (in units of the boson charge)
is bound at an interface between the BTI state and the vacuum (or a trivial state). 
One possible model for the bulk of the BTI state is an $O(3)$ NLSM with theta term and coefficient $\theta=2\pi k$, 
$k\in\mathbb{Z}$~\cite{CenkeClass1}. The boundary of the BTI state is then described by the same NLSM but with a WZ term
at level $k$. In $1+1$ dimensions SPT phases with the symmetry group $G= U(1)\rtimes\mathbb{Z}_2$ have a 
$\mathbb{Z}_2$ classification, meaning that there is only a single nontrivial phase~\cite{chen2013symmetry,CenkeClass1}. 
This single nontrivial phase is the BTI state. Within the NLSM description, the NLSM for any odd $k$ represents the nontrivial BTI 
state, while the model for any even $k$ represents the trivial state.

In the $O(3)$ NLSM description the field is a unit vector field $\mb{n}$ with components $n^a$, $a=1,2,3$. The target space of the
$O(3)$ NLSM is the unit two-sphere $S^2$. As in Sec.~\ref{sec:pert}, the action of the symmetry group 
$G= U(1)\rtimes\mathbb{Z}_2$ of the BTI on the NLSM field is best expressed by first combining $n^1$ and $n^2$ into the 
``boson" field $b= n^1 +i n^2$. Then for the BTI state, the action of $G$ on the NLSM field is given by 
(see Sec.~IV.D of Ref.~\onlinecite{CenkeClass1})
\beq
	U(1): b\to e^{i\xi} b\ , \label{eq:U1}
\eeq
and
\bseq
\label{eq:Z2-action}
\beqa
	\mathbb{Z}_2: b &\to& b^* \\
	n^{3} &\to& -n^{3}\  .
\eeqa
\eseq
Since the $\mathbb{Z}_2$ symmetry is unitary, the transformation $b\to b^*$ is equivalent to $n^1\to n^1$, $n^2\to -n^2$. 
We can interpret $b$ as the field which annihilates a boson of charge $1$, and
$n^3$ can be interpreted as the deviation of the boson density from a non-zero constant value. 

The theta term and the WZ term for the $O(3)$ NLSM are both expressed in terms of the volume form $\omega_2$ on $S^2$, 
\beq
	\omega_{2}= n^1 dn^2 \wedge dn^3 - n^2 dn^1\wedge dn^3 + n^3 dn^1\wedge dn^2\ . \label{eq:volume-form}
\eeq
In what follows we use $\A_{2}=4\pi$ to denote the surface area of $S^2$ (and $\int_{S^2}\omega_{2}= \A_{2}$).
In this article we are only interested in the boundary theory of the BTI, and so we focus our attention on the WZ term. 
The boundary theory lives in one spacetime dimension. To make our discussion 
as precise as possible, we take the time coordinate (the only coordinate here) to lie in the interval $t\in[0,T)$, and we impose
periodic boundary conditions in the time direction. This makes our one-dimensional spacetime into a circle of circumference $T$.
Let us denote the one-dimensional spacetime by $S^1_T$ (the circle of circumference $T$). 
Constructing the WZ term requires extending the spacetime
into a two-dimensional spacetime $\mathcal{B}$ such that $\pd \mathcal{B}= S^1_T$. We use $\tilde{\mb{n}}$ to denote the
extension of the NLSM field $\mb{n}$ into the bulk of $\mathcal{B}$, and we require that
$\tilde{\mb{n}}|_{\pd\mathcal{B}}=\mb{n}$. Using $\mathcal{B}$ and the extension $\tilde{\mb{n}}$ of
$\mb{n}$, the WZ term takes the form
\beq
	S_{WZ}[\mb{n}]= \frac{2\pi k}{\A_2}\int_{\mathcal{B}}\tilde{\mb{n}}^*\omega_2\ ,
\eeq
where $\tilde{\mb{n}}^*\omega_2$ denotes the pullback of $\omega_2$ to $\mathcal{B}$ via the map 
$\tilde{\mb{n}}: \mathcal{B} \to S^2$, and $k$ is the level of the WZ term (the same integer $k$ determines
the coefficient $\theta=2\pi k$ of the theta term describing the bulk of the SPT phase).

The complete $O(3)$ NLSM action describing the boundary of the BTI takes the form
\begin{align}
	S_{bdy}[\mb{n}] = \int_0^T &dt\ \frac{1}{2 f_{bdy}}\left[(\pd^{t}b)^*(\pd_t b) + (\pd^t n^3)(\pd_t n^3)\right] \nnb \\
+\ &S_{WZ}[\mb{n}]\ ,
\end{align}
where $f_{bdy}$ is a boundary coupling constant and $\pd^t= \pd_t$ for our choice of the signature of the spacetime metric 
(we use a ``mostly minus" Minkowski metric). We can now consider coupling the boundary theory to an external $U(1)$
gauge field $A=A_t dt$. In Ref.~\onlinecite{lapa2017}, we showed that the properly gauged boundary action has the form
\begin{widetext}
\beq
	S_{bdy,gauged}[\mb{n},A]= \int_0^T dt\ \frac{1}{2 f_{bdy}}\left[(D^{t}b)^*(D_t b) + (\pd^t n^3)(\pd_t n^3)\right] + S_{WZ,gauged}[\mb{n},A]\ ,
\eeq
\end{widetext}
where
\beq
	S_{WZ,gauged}[\mb{n},A]= S_{WZ}[\mb{n}] + \frac{2\pi k}{\A_2}\int_0^T dt\  n^3 A_t\ , \label{eq:O3-gauged-WZ}
\eeq
and $D_t= \pd_t - i A_t$ ($\pd_t = \pd^t$, $A_t= A^t$, etc., for our choice of signature). 
The action for the fully gauged boundary theory is invariant under $U(1)$ gauge transformations
\begin{subequations}
\beqa
	b &\to& e^{i\xi} b \\
	A &\to& A + d\xi\ ,
\eeqa
\end{subequations}
and $\mathbb{Z}_2$ transformations
\begin{subequations}
\beqa
	b &\to& b^* \\
	n^{3} &\to& -n^{3} \\
	A &\to& -A\ .
\eeqa
\end{subequations}

\subsection{Boundary partition function and global anomaly}

We now study the partition function for the gauged boundary theory of the BTI in the topological limit
$f_{bdy}\to\infty$. In this limit we keep only the low energy information about the boundary theory, including possible
anomalies. The partition function 
\beq
	Z[A]= \int [d\mb{n}]\ e^{iS_{bdy,gauged}[\mb{n},A]}\ 
\eeq
of the gauged boundary theory can be evaluated very simply in this limit, as we now discuss. 
First, in the limit $f_{bdy}\to \infty$ the path integral we need to evaluate is
\beq
	Z[A]= \int [d\mb{n}]\ e^{iS_{WZ,gauged}[\mb{n},A]}\ , \label{eq:BTI-bdy-partition-fcn}
\eeq
where $S_{WZ,gauged}[\mb{n},A]$ is the gauged WZ action from Eq.~\eqref{eq:O3-gauged-WZ}.
The path integral measure appearing here has the precise definition
\beq
	 [d\mb{n}]= \prod_{t\in[0,T)}\omega_2(t)\ , \label{eq:measure-O3-integral}
\eeq
where $\omega_2(t)$ denotes the volume form for a copy of $S^2$ located at the point $t$ in spacetime, and we integrate 
over all field configurations with periodic boundary conditions in time.

We can also use a gauge transformation to simplify the form of the coupling to the gauge field $A$. In one spacetime dimension
the gauge field one-form $A= A_t dt$ has only one component. Since our spacetime is a circle, which has first cohomology
group $H^1(S^1,\mathbb{R})= \mathbb{R}$, we can decompose a generic $A_t$ as
\beq
	A_t = \overline{A}_t + \pd_t \lambda\ ,
\eeq
where 
\beq
	\overline{A}_t := \frac{1}{T}\int_0^T dt\ A_t\ 
\eeq
represents the nontrivial part of $A$, and $\pd_t\lambda$ represents the exact part of $A$ (here $\lambda$ is some function of 
$t$). The exact part of $A$ can be removed from the action via a small $U(1)$ gauge transformation, which are those gauge 
transformations $A\to A + \pd_t\xi$ with the function $\xi$ satisfying $\xi(0)=\xi(T)$.  Large $U(1)$ gauge transformations are 
those transformations which send $\overline{A}_t \to \overline{A}_t + \frac{2\pi n}{T}$, for any $n\in\mathbb{Z}$, and they will 
play an important role in the discussion of the global anomaly in this theory later in this section. The upshot of all of this is that we can 
replace the coupling to $A_t$ in the gauged WZ action with a coupling to the \emph{constant} gauge field $\overline{A}_t$.

We now move on to the calculation of the partition function $Z[A]$. We compute the partition function by observing that it is 
identical to the phase space path integral for a spin of magnitude $J= \frac{k}{2}$ (or $\frac{|k|}{2}$ for negative $k$) in a constant 
magnetic field $B$ pointing in the 3-direction, with the magnitude of the magnetic field given in terms of the gauge field $A$ by 
$B= -\overline{A}_t$. To prove this we now briefly review the form of the phase space path integral for spin. At this point we 
recommend that the reader skim through Appendix~\ref{app:symp-geom} where we review the phase space path integral 
expression for the partition function of a quantum mechanical system obtained by quantizing a general classical system defined on 
a phase space $\M$ equipped with a symplectic form $\omega$ and Hamiltonian function $H$. 

The classical mechanics of a spin $J= \frac{1}{2},1,\frac{3}{2},\dots$,  is described by a phase space $\mathcal{M}=S^2$ 
equipped with the symplectic form $\omega = J\omega_2$, where $\omega_2$ is the volume form on $S^2$
from Eq.~\eqref{eq:volume-form}. It is convenient to work in spherical coordinates $(\phi,\theta)$ on $S^2$. In this system
of coordinates the components of the NLSM field $\mb{n}$ take the form
\begin{subequations}
\beqa
	n^1 &=& \sin(\theta)\cos(\phi) \\
	n^2 &=& \sin(\theta)\sin(\phi) \\
	n^3 &=& \cos(\theta)\ ,
\eeqa
\end{subequations}
and we have
\beq
	\omega= J\sin(\theta) d\theta\wedge d\phi\ .
\eeq
Using the definition Eq.~\eqref{eq:Poisson-bracket} for the Poisson bracket one can check that 
\beq
	\{n^a,n^b\}= \frac{1}{J}\sum_c \ep^{abc} n^c\ ,
\eeq
so that the spin components $S^a$ are given in terms of $n^a$ by
\beq
	S^a= J n^a\ .
\eeq
The spin components then obey the Poisson algebra
\beq
	\{S^a,S^b\}= \sum_c \ep^{abc} S^c\ .
\eeq
We can now see that replacing the Poisson bracket with a commutator according to the rule $\{\cdot,\cdot\} \to -i[\cdot,\cdot]$
will give the usual commutation relations for spin in quantum mechanics.

Now let us assume that the dynamics of the spin system are specified by a Hamiltonian $H$. Then the phase space path integral 
representing the partition function $\text{tr}_J[e^{-iHT}]$, where the trace is taken in the spin $J$ representation
of $SU(2)$, takes the form
\beq
	\text{tr}_J[e^{-iHT}]= \int [d\phi d\theta]\left[ \prod_{t\in[0,T)} J \sin(\theta(t))\right] e^{iS[\phi,\theta]}\ .
\eeq
where
\beq
	S[\phi,\theta]= \int_0^T dt\ \left[ \vth_{\phi}\pd_t \phi + \vth_{\theta}\pd_t\theta- H(\theta,\phi) \right]\ . 
\eeq
Here $\vth_{\phi}$ and $\vth_{\theta}$ are the components of the symplectic potential $\vth$, which is defined locally
on the phase space by the relation $\omega= d\vth$ (Eq.~\eqref{eq:symp-potential} in Appendix~\ref{app:symp-geom}). 
Then, since
\beq
	(\vth_{\phi}\pd_t \phi + \vth_{\theta}\pd_t\theta)dt = \mb{n}^*\vth\ ,
\eeq
we can rewrite the first term in this action using an extension $\mathcal{B}$ of the spacetime $S^1_T$ and an
extension $\tilde{\mb{n}}$ of the field configuration (satisfying $\tilde{\mb{n}}|_{\pd\B}=\mb{n}$). We have
\beqa
	\int_{S^1_T}\mb{n}^*\vth &=& \int_{\mathcal{B}}\tilde{\mb{n}}^*\omega \nnb \\
	&=& J\int_{\mathcal{B}}\tilde{\mb{n}}^*\omega_2 \ ,
\eeqa
where the first line follows from Stokes' theorem. If we choose the Hamiltonian to be
\beq
	H= B S^3= B J n^3\ ,
\eeq
which is the Hamiltonian for a spin in a constant magnetic field of magnitude $B$ and pointing in the 3-direction,
then the action becomes
\beq
	S[\phi,\theta]= J\int_{\mathcal{B}}\tilde{\mb{n}}^*\omega_2 - J\int_{S^1_T}n^3 B\ .
\eeq

We can now compare the path integral for a spin in a magnetic field to our path integral in Eq.~\eqref{eq:BTI-bdy-partition-fcn} 
for the partition function of the boundary of the BTI state. Using the fact that $\A_2= 4\pi$, we find that these path integrals are 
identical if we make the identifications
\begin{subequations}
\beqa
	J &=&\frac{k}{2} \\
	B &=& -\overline{A}_t\ .
\eeqa
\end{subequations}
More precisely, the path integrals are not identical but differ by the infinite constant factor
\beq
	\prod_{t\in[0,T)} J\ ,
\eeq
but we can give a more careful definition of the path integral measure for the partition function of the
gauged boundary theory for the BTI by including this factor. Using all of this information we then find that
\beqa
	Z[A] &=& \text{tr}_{\frac{k}{2}}[e^{i S^3 \overline{A}_t T}] \nnb \\
		&=& \sum_{j=-\frac{k}{2}}^{\frac{k}{2}} e^{i j\overline{A}_t T} \nnb \\
		&=& \frac{\sin\left[\frac{\overline{A}_t T}{2}(k+1)\right]}{\sin\left[\frac{\overline{A}_t T}{2}\right]}\ . \label{eq:bdy-effective-action}
\eeqa
Note that in deriving this formula we assumed that $k>0$. For $k<0$ one just needs to replace $k$ with $|k|$. For the 
discussion below it is useful to decompose the gauge field as $\overline{A}_t= \frac{2\pi \ell}{T} + \overline{a}_t$ for some 
$\ell\in\mathbb{Z}$ and $\overline{a}_t \in (0,\frac{2\pi}{T})$, and to then rewrite $Z[A]$ in terms of $\ell$ and $\overline{a}_t$,
\beq
	Z[A] = (-1)^{k\ell}\frac{\sin\left[\frac{\overline{a}_t T}{2}(k+1)\right]}{\sin\left[\frac{\overline{a}_t T}{2}\right]}\ .
\eeq
It is important to observe that the factor $(-1)^{k\ell}$ is nontrivial for odd $k$. This minus sign is related to the global
anomaly in this theory for odd $k$, as we now discuss. 

For any level $k$ the partition function $Z[A]$ respects the $\mathbb{Z}_2$ symmetry of the BTI state, i.e., we have
\beq
	Z[-A]= Z[A]\ .
\eeq
However, for odd $k$ the partition function is not invariant under a large $U(1)$ gauge transformation,
\beq
	\overline{A}_t \to \overline{A}_t + \frac{2\pi}{T}\ ,
\eeq
which is equivalent to the transformation $\ell \to \ell + 1$ if we decompose the gauge field as 
$\overline{A}_t= \frac{2\pi \ell}{T} + \overline{a}_t$. Instead, for odd $k$ the partition function $Z[A]$ 
changes sign under this transformation. We can try to fix this large gauge invariance issue by modifying the partition function to 
\beq
	\tilde{Z}[A]= Z[A]e^{\pm \frac{i}{2}\overline{A}_t T}\ .
\eeq
This is equivalent to adding the local counterterm $\pm \frac{1}{2} \int_0^T dt\  A_t$ to the original boundary action,
which is a $(0+1)$-dimensional Chern-Simons term with fractional level $\pm\frac{1}{2}$. Note, 
however, that adding this counterterm spoils the invariance of the partition function under the action of the $\mathbb{Z}_2$ 
symmetry. Therefore we find that although the gauged action $S_{WZ,gauged}[\mb{n},A]$ for the BTI boundary has large 
$U(1)$ gauge invariance \emph{and} $\mathbb{Z}_2$ symmetry, the partition function $Z[A]$ for the boundary theory only has
both of these symmetries when $k$ is even. 

This is a classic sign of a \emph{global anomaly} in the $\mathbb{Z}_2$
symmetry: for odd $k$ we can quantize the theory in such a way as to keep either the $\mathbb{Z}_2$ symmetry or  
large $U(1)$ gauge invariance, but not both. Physically, this anomaly is related to the fact that for odd $k$ the boundary
of the BTI has states with half-integer (i.e., fractional) charge. In addition, the fact that the presence or absence of the anomaly 
depends only on the parity of $k$ (even or odd) is due to the aforementioned $\mathbb{Z}_2$ classification of 
bosonic SPT phases with $G=U(1)\rtimes\mathbb{Z}_2$ symmetry in $1+1$ dimensions (the theories with odd $k$ all represent 
the nontrivial BTI state, while the theories with even $k$ all represent the trivial phase). As we discussed above,
the anomaly here is very similar to the global anomaly computed in Ref.~\onlinecite{elitzur1986origins} for a Dirac fermion in 
$0+1$ dimensions with $U(1)$ and $\mathbb{Z}_2$ symmetry. In addition, a similar anomaly in the (purely
bosonic) $(0+1)$-dimensional theory of a particle on a ring was discussed recently in Appendix~D of 
Ref.~\onlinecite{gaiotto2017theta}.

\subsection{Deforming the target space}

In the previous subsection we showed that, at least within the $O(3)$ NLSM description, the boundary of the $(1+1)$-dimensional
 BTI phase exhibits a global anomaly in the $\mathbb{Z}_2$ symmetry of the BTI phase. However, our derivation of the anomaly 
seemed to rely on the specific geometry of the target space $S^2$ of the $O(3)$ NLSM. Specifically, our derivation used the
fact that the partition function for the BTI boundary was equivalent to a phase space path integral for a spin in a magnetic field. 
In addition, since $U(1)\rtimes \mathbb{Z}_2$ is a subgroup of $SO(3)$, the anomaly we derived is closely related to the global 
$SO(3)$ anomaly of the $O(3)$ NLSM with WZ term in $0+1$ dimensions 
(see, for example, the discussion in Sec.~1.2 of Ref.~\onlinecite{benini2017comments}).
Our calculation then shows that the $U(1)\rtimes\mathbb{Z}_2$ subgroup of $SO(3)$ is also anomalous in this theory.

In the rest of this section we show that the boundary anomaly of the BTI
state is not affected by any smooth deformation of the target space $S^2$ of the $O(3)$ NLSM which also preserves the 
$U(1)\rtimes \mathbb{Z}_2$ symmetry of the BTI phase. In other words, we break the $SO(3)$ symmetry of the model
down to $U(1)\rtimes\mathbb{Z}_2$, and we show that the anomaly still exists in these less symmetric theories.

In this subsection we describe the geometry of such deformed target spaces, 
and then we construct models of the BTI boundary using WZ terms for NLSMs with these deformed 
target spaces. We also show how to properly gauge these WZ actions. In the next subsection we use the 
equivariant localization (EL) technique to compute the partition function for these models, and we show that all such 
models have a partition function which is identical to Eq.~\eqref{eq:bdy-effective-action}. Thus, we find that
the boundary anomaly is completely unaffected by smooth, symmetry-preserving deformations of the target 
space of the NLSM.

As stated above, we consider descriptions of the BTI using NLSMs with a target space $\M$ that can be obtained from the target 
space $S^2$ of the $O(3)$ NLSM by smooth deformations which preserve the $G= U(1)\rtimes \mathbb{Z}_2$ symmetry of the 
BTI phase. As in Sec.~\ref{sec:pert}, we can characterize such spaces $\M$ precisely through the notion of a diffeomorphism which
is \emph{equivariant} with respect to the symmetry of the BTI phase. The target space of the $O(3)$ NLSM is $S^2$, and the
NLSM description of the BTI phase includes an action of the group $G= U(1)\rtimes\mathbb{Z}_2$ on $S^2$. This action was
shown explicitly in Eq.~\eqref{eq:U1} and Eqs.~\eqref{eq:Z2-action}.
Let us assume that
the manifold $\M$ is also equipped with an action of the group $G$. Then a diffeomorphism $f:\M \to S^2$ is equivariant with respect 
to $G$ if 
\beq
	f(g\cdot \mb{m})= g\cdot f(\mb{m}),\ \forall g\in G\ ,\ \forall \mb{m}\in \M\ .
\eeq
This is the correct mathematical notion corresponding to the intuitive idea of a manifold which can be obtained from $S^2$ by
smooth, symmetry-preserving deformations. 

The spaces $\M$ which are related to $S^2$ in this way
 can be realized as surfaces of revolution in $\mathbb{R}^3$ which are symmetric under rotation about
the $z$-axis (this guarantees $U(1)$ symmetry), and which are also invariant under reflection $z\to-z$ through the
$x$-$y$ plane\footnote{We use standard Cartesian coordinates $x,y,z$ for $\mathbb{R}^3$.}.
The latter condition guarantees that $\M$ possesses the $\mathbb{Z}_2$ symmetry of the BTI phase. These spaces
$\M$ are completely specified by a parametric curve $(r(\sigma),z(\sigma))$, where $r(\sigma)$ is the distance of the
surface from the $z$-axis in $\mathbb{R}^3$ at the height $z(\sigma)$, and $\sigma\in[a,b]$ is a parameter used to specify the
curve. If we think of $(r(\sigma),z(\sigma))$ as, say, a curve in the $x$-$z$ plane (replace $r$ with $x$), then we can imagine
constructing the full surface $\M$ by rotating the curve about the $z$-axis in $\mathbb{R}^3$. We can then choose
coordinates on $\M$ to be $(\sigma,\phi)$, where $\phi$ is the usual azimuthal angle in spherical or cylindrical coordinates
in $\mathbb{R}^3$. Finally, in order for this construction to produce a smooth manifold (with no conical singularities at the
top and bottom), we require that $\frac{dz}{dr}=0$ at the top and bottom of the curve. This is equivalent to the
condition
\beq
	\frac{\pd_{\sigma}z(\sigma)|_{\sigma=a,b}}{\pd_{\sigma}r(\sigma)|_{\sigma=a,b}} = 0\ ,
\eeq
or just
\beq
	\pd_{\sigma}z(\sigma)|_{\sigma=a,b}= 0\ ,
\eeq
assuming that $\pd_{\sigma}r(\sigma)$ does not vanish at $\sigma=a,b$.
 
In principle we can use any parametrization of the surface, but the most convenient choice is a parametrization $(r(s),z(s))$ in 
terms of the arc length $s$ along the curve, where 
\beq
	s(\sigma)= \int_a^{\sigma} d\sigma' \sqrt{\left(\pd_{\sigma'}r(\sigma')\right)^2 +  \left(\pd_{\sigma'}z(\sigma')\right)^2}\ .
\eeq
We define $L= s(b)$ to be the total length of the curve. In the coordinate system $(s,\phi)$, the metric on $\M$ takes the
form
\beq
	g= ds\otimes ds + [r(s)]^2 d\phi\otimes d\phi\ ,
\eeq
and the volume form is
\beq
	\omega_{\M}= r(s) ds\wedge d\phi\ .
\eeq
The total area of the target space is then $\A_{\M}= 2\pi \int_0^L ds\ r(s)$. In addition, the ``unit speed'' property 
$(\pd_s r(s))^2 + (\pd_s z(s))^2=1$ of the arc length parametrization, combined with the restriction 
$\pd_s z(s)|_{s=0,L}=0$, implies that $\pd_s r(s)|_{s=0}= 1$ and $\pd_s r(s)|_{s=L}= -1$. The signs here follow
from the fact that the width of the surface $\M$ \emph{increases} from zero near $s=0$ and \emph{decreases} back to zero at 
$s=L$. We also assume that $z(s)= -z(L-s)$ so that $\M$ is symmetric under reflection through the $z=0$ plane in 
$\mathbb{R}^3$.

We can now construct a model for the boundary of the BTI using the NLSM with target space $\M$. We denote the 
NLSM field by $\mb{m}= (m^1,m^2)$, with components $m^1= s$ and $m^2= \phi$. In the low energy (topological) 
limit the boundary action contains only a WZ term for $\mb{m}$. As usual, to construct this term we require an extension 
$\mathcal{B}$ of the boundary spacetime $S^1_T$, and an extension $\tilde{\mb{m}}$ of the NLSM field $\mb{m}$
 into the bulk of $\B$. Then the WZ action describing the low energy physics of the boundary is
\beq
	S_{WZ}[\mb{m}]= \frac{2\pi k}{\A_{\M}}\int_{\B}\tilde{\mb{m}}^*\omega_{\M}\ ,
\eeq
where $k\in\mathbb{Z}$ is the level of the WZ term. We choose the $U(1)$ and $\mathbb{Z}_2$ symmetries of the BTI state 
to act on the components of the field $\mb{m}$ as
\beq
	U(1): \phi \to \phi +\xi\ ,
\eeq
and
\bseq
\beqa
	\mathbb{Z}_2: \phi &\to& -\phi \\
	s &\to& L-s\ .
\eeqa
\eseq
This action of the $\mathbb{Z}_2$ symmetry is the generalization to the target space $\M$ of the $\mathbb{Z}_2$ action on
$S^2$ from Eqs.~\eqref{eq:Z2-action}. 

The next step is to gauge the $U(1)$ symmetry by coupling the boundary WZ action to the gauge 
field $A= A_t dt$. One can check that the action 
\beq
	S_{WZ,gauged}[\mb{m},A]= S_{WZ}[\mb{m}] - \frac{2\pi k}{\A_{\M}}\int_0^T dt\ f(s(t)) A_t\ \label{eq:WZ-gauged-general-target}
\eeq
will be invariant under the gauge transformation $\phi\to \phi +\xi$, $A\to A+d\xi$, if the function $f(s)$ satisfies the
first order differential equation
\beq
	\pd_s f(s)= r(s)\ .
\eeq
This equation has the simple solution $f(s)= C+ \int_0^s ds'\ r(s')$, where $C$ is an as yet undetermined constant. However,
since we require the gauged action to be invariant under the charge-conjugation operation
\bseq
\beqa
	\mathbb{Z}_2: \phi &\to& -\phi \\
	s &\to& L-s \\
	A &\to& -A\ ,
\eeqa
\eseq
we find that this constant is fixed to take the value $C=-\frac{\A_{\M}}{4\pi}$. Therefore the function $f(s)$ appearing
in the gauged boundary action is given by
\beq
	f(s)= \int_0^s ds'\ r(s') - \frac{\A_{\M}}{4\pi}\ .
\eeq
In particular we have 
\beq
	f(L)= -f(0)= \frac{\A_{\M}}{4\pi}\ ,
\eeq
which will be needed for the calculation of the partition function in the next subsection.

\subsection{Boundary partition function and global anomaly for all target spaces}

We now turn to the evaluation of the partition function $Z[A]$ for the 
NLSM with target space $\M$ and action given by Eq.~\eqref{eq:WZ-gauged-general-target} using the equivariant localization 
(EL) technique.
We give a brief introduction to the EL technique in Appendix~\ref{app:EL}, and in 
Appendix~\ref{app:dets} we show how to calculate the Pfaffians which appear in the final expression for $Z[A]$. 
Therefore, in this section we only outline the calculation and present the result. 
The final result for the partition function turns out to be \emph{completely identical} to the partition function 
of Eq.~\eqref{eq:bdy-effective-action} which we derived for the special case of the $O(3)$ NLSM with target space $S^2$. 
The mechanism which underlies the EL technique allows us to understand why this is the case. First, the EL technique applied
to our particular problem yields the result that the partition function depends only on field configurations $\mb{m}$
near the points on $\M$ which are fixed by the $U(1)$ action. These are just the two
points $s=0$ and $s=L$ at the bottom and the top of $\M$. The value of the gauged WZ action at these two points is actually
\emph{independent} of the specific choice of the target space $\M$ (see Eqs.~\eqref{eq:action-at-fixed-points} below). 
Therefore we find that since the partition function only receives contributions from
field configurations near $s=0$ and $s=L$, and since the action at those two points is independent of the details of $\M$, the
partition function $Z[A]$ is independent of the specific details of the target space $\M$. The discussion here is meant
to be heuristic, and so we now move on to a more detailed presentation of the calculation.

We start by rewriting the gauged WZ action for the NLSM in a way which makes the problem of computing the partition function
of this theory look like a phase space path integral for a dynamical system with phase space $\M$. The reason for this is
that the EL technique, in its original formulation, applies to phase space path integrals. To achieve this goal we
first recall that we can use a small $U(1)$ gauge transformation to replace the gauge field $A_t$ with its time average 
$\overline{A}_t$ in the gauged WZ action. Next, we rewrite the gauged WZ action as
\beq
	S_{WZ,gauged}[\mb{m},A]= \int_{\mathcal{B}}\tilde{\mb{m}}^*\omega - \int_0^T dt\ H(\mb{m}) \ ,
\eeq
where we defined
\begin{subequations}
\label{eq:symp-form-and-Ham}
\beqa
	\omega &=& \frac{2\pi k}{\A_{\M}}\omega_{\M} \\
	H(\mb{m}) &=& \frac{2\pi k}{\A_{\M}} f(s) \overline{A}_t\ .
\eeqa
\end{subequations}

We can now see that the path integral for $Z[A]$ is equivalent to a phase space path integral 
(see our Appendix~\ref{app:symp-geom} for a review) 
for a dynamical system described by the triple $(\M,\omega,H)$, with the symplectic form $\omega$ and Hamiltonian $H$ 
defined by Eqs.~\eqref{eq:symp-form-and-Ham}. The Hamiltonian $H$ and the symplectic form $\omega$ are related via the 
equation $dH= -i_{\underline{v}}\omega$, where the vector field $\un{v}$ is given by
\beq
	\un{v}= \overline{A}_t\pd_{\phi}\ . \label{eq:hamiltonian-vector-field}
\eeq
This vector field is clearly proportional to the vector field $\pd_{\phi}$ which generates the action of
the $U(1)$ part of the symmetry group $G=U(1)\rtimes\mathbb{Z}_2$ of the BTI on the target space $\M$ of the NLSM. 
The classical equations of motion for this system are
\bseq
\beqa
	\dot{s} &=& 0 \\
	\dot{\phi} &=& \overline{A}_t\ .
\eeqa
\eseq
These equations say that (classically) each point on $\M$ revolves around the $z$-axis in $\mathbb{R}^3$ with a period 
$\frac{2\pi}{\overline{A}_t}$. In the notation of Appendix~\ref{app:EL} the classical equations of motion can be rewritten as
$V^a_S[\mb{m}(t);t]= 0$, $a=1,2$, where $V^a_S[\mb{m}(t);t]= \dot{m}^a(t) - v^a(\mb{m}(t))$ and
$v^a$ are the components of the vector field $\un{v}$ from Eq.~\eqref{eq:hamiltonian-vector-field}.

We are now almost ready to to apply the EL results from Appendix~\ref{app:EL} to compute the partition function. First, let us 
assume that $T \neq \frac{2\pi n}{\overline{A}_t}$ for any $n\in\mathbb{Z}$. This means that the only $T$-periodic solutions to 
the classical equations of motion for the dynamical system defined by $(\M,\omega,H)$ are the \emph{constant} solutions $s=0$ 
and $s=L$. Therefore, the set $L\mathcal{M}_S$ of $T$-periodic solutions to the classical equations of motion 
(defined in Eq.~\eqref{eq:set-of-periodic-solutions}) has only these 
two elements, and the final result for the partition function $Z[A]$ only involves contributions from field
configurations close to these solutions. Using the EL technique we find that the partition function can be expressed
only in terms of contributions from $s=0$ and $s=L$ as
\begin{align}
	Z[A] \sim \frac{e^{iS_{WZ,gauged}[\mb{m},A]_{s=0}}}{\text{Pf}[\mathcal{O}]_{s=0}} + \frac{e^{iS_{WZ,gauged}[\mb{m},A]_{s=L}}}{\text{Pf}[\mathcal{O}]_{s=L}}\ , \label{eq:EL-formula-for-Z}
\end{align}
where the operator $\mathcal{O}$ is defined in Eq.~\eqref{eq:operator-O} of Appendix~\ref{app:EL}.
The value of the gauged WZ action at these two solutions is
\bseq
\label{eq:action-at-fixed-points}
\beqa
	S_{WZ,gauged}[\mb{m},A]_{s=0} &=& \frac{k}{2}\overline{A}_t T \\
	S_{WZ,gauged}[\mb{m},A]_{s=L} &=& -\frac{k}{2}\overline{A}_t T\ .
\eeqa
\eseq
Remarkably, these expressions do not depend on the area $\A_{\M},$ or any other details, of the target space $\M$.
We now turn to the evaluation of the Pfaffians appearing in the denominators in Eq.~\eqref{eq:EL-formula-for-Z}.

To calculate the Pfaffians (which by Eq.~\eqref{eq:operator-O} depend on the derivatives of the vector field $\un{v}$), 
we first need to express $\un{v}$ in a system of local coordinates $(x,y)$ near
the points $s=0$ and $s=L$ of the space $\M$. The coordinate system $(s,\phi)$ is singular at these two points
($\phi$ is undefined there) and so it cannot be used for an analysis of the space near these two points. Near $s=0$ we 
choose coordinates $x= \frac{2\pi k}{\A_{\M}}s\cos(\phi)$, $y=\frac{2\pi k}{\A_{\M}}s \sin(\phi)$, and near $s=L$ we choose 
coordinates $x= -\frac{2\pi k}{\A_{\M}}(L-s)\cos(\phi)$, $y= \frac{2\pi k}{\A_{\M}}(L-s)\sin(\phi)$. 
This choice of coordinates has the virtue that the symplectic form $\omega$ takes the Darboux form $\omega= dx\wedge dy$ at 
both $s=0$ and $s=L$. To derive this result we had to use the important property that 
$\pd_s r(s)|_{s=0} = -\pd_s r(s)|_{s=L}= 1$. For these choices of coordinates the vector field $\un{v}$ takes the form
\beq
	\un{v}= \overline{A}_t(x \pd_y - y\pd_x)
\eeq
near $s=0,$ and the form
\beq
	\un{v}= -\overline{A}_t(x \pd_y - y\pd_x)
\eeq
near $s=L$.

Using the definition of $\text{Pf}[\mathcal{O}]$ in terms of a fermion path integral from 
Eq.~\eqref{eq:Pfaffian-as-path-integral} of Appendix~\ref{app:EL}, 
we find that $\text{Pf}[\mathcal{O}]$ at the points $s=0$ and $s=L$ is given formally
by the determinant of a one-dimensional Dirac operator. More precisely, after expanding the path integral in Fourier
modes we find that
\beqa
	\text{Pf}[\mathcal{O}]_{s=0} &=& -\overline{A}_t \prod_{m>0}\left( \frac{2\pi m}{T} +  \overline{A}_t \right)\left( -\frac{2\pi m}{T} +  \overline{A}_t \right) \nnb \\ 
	&=& -\text{det}[-i\pd_t + \overline{A}_t]\ ,
\eeqa
and
\beq
	\text{Pf}[\mathcal{O}]_{s=L}=  -\text{det}[-i\pd_t - \overline{A}_t]\ .
\eeq
The operators 
\beq
	\mathcal{D}_{\pm} := -i\pd_t \pm \overline{A}_t
\eeq
are equivalent to one-dimensional Dirac operators for a fermion in one spacetime dimension coupled to the external
field $A= A_t dt$~\cite{elitzur1986origins}. 
As we discussed in Appendix~\ref{app:EL}, the overall sign of these Pfaffians is ambiguous, since 
we are free to alter the order of factors in the definition of the path integral measure. Therefore at this point we are free to 
choose a particular definition of the path integral measure such that 
\beqa
	\text{Pf}[\mathcal{O}]_{s=0} &=& \text{det}[\mathcal{D}_{+} ] \\
	\text{Pf}[\mathcal{O}]_{s=L} &=& \text{det}[\mathcal{D}_{-} ]\ .
\eeqa
These determinants still require proper regularization, and we now turn to a discussion of this issue.

We choose to regularize these determinants using \emph{zeta} and \emph{eta} function methods
(see Appendix~\ref{app:dets} for details). To motivate the definition of the
regularized determinants in terms of zeta and eta functions, we first consider the following (non-rigorous) manipulations
of a definition of these determinants in terms of an infinite product of their eigenvalues. We are also careful to point out
any ambiguities which arise in defining the determinants in this way. Let 
$\lam^{(\pm)}_m=  \frac{2\pi m}{T} \pm \overline{A}_t$, $m\in\mathbb{Z}$,
 be the eigenvalues of the operator $\mathcal{D}_{\pm}$. Formally, we have
\beqa
	\text{det}[\mathcal{D}_{\pm}] &=& \prod_{m\in\mathbb{Z}} \lam^{(\pm)}_m \nnb \\
	&=& \prod_{m\in\mathbb{Z}} |\lam^{(\pm)}_m| \text{sgn}(\lam^{(\pm)}_m)\ .
\eeqa
So far we encounter no difficulties. However, the next step is to express the sign of the eigenvalues as
\beq
	\text{sgn}(\lam^{(\pm)}_m) = e^{i\frac{\pi}{2}(1-\text{sgn}(\lam^{(\pm)}_m))}\ .
\eeq
But this step is ambiguous because we could just as well have written 
\beq
	\text{sgn}(\lam^{(\pm)}_m) = e^{i\frac{(2p+1)\pi}{2}(1-\text{sgn}(\lam^{(\pm)}_m))}
\eeq
for \emph{any} integer $p$. 
For now we work with the most general expression for $\text{sgn}(\lam^{(\pm)}_m)$, which involves an arbitrary integer $p$.
Later in this section we show how the value of $p$ can be fixed by a minimal number of physical assumptions on the properties
of the partition function $Z[A]$. 

Continuing with our manipulations, we find that the determinant can be expressed formally as
\begin{align}
	\text{det}[\mathcal{D}_{\pm}]= \left(\prod_{m\in\mathbb{Z}} |\lam^{(\pm)}_m| \right)e^{i\frac{(2p+1)\pi}{2}\sum_{m\in\mathbb{Z}}(1-\text{sgn}(\lam^{(\pm)}_m))}\ .
\end{align}
We now use zeta and eta function methods to make sense of the different terms in this expression. Before we start,
we again decompose $\overline{A}_t$ as $\overline{A}_t= \frac{2\pi \ell}{T} + \overline{a}_t$, for some $\ell\in\mathbb{Z}$
and $\overline{a}_t\in(0,\frac{2\pi}{T})$. To start with the regularization, we first use
zeta function regularization to define the product over the magnitude of all the eigenvalues $\lam^{(\pm)}_m$. 
We carry out this calculation in Appendix~\ref{app:dets} and we find that
\beq
	\left(\prod_{m\in\mathbb{Z}} |\lambda_m| \right)_{reg}= 2\sin\left( \frac{\overline{a}_t T}{2}\right)\ .
\eeq
Next, we define the sum $\sum_{m\in\mathbb{Z}} 1$ as
\beq
	\left(\sum_{m\in\mathbb{Z}} 1 \right)_{reg} = 1 + 2\zeta(0) = 0\ ,
\eeq
where $\zeta(s)= \sum_{n=1}^{\infty} \frac{1}{n^s}$ is the Riemann zeta function and we used $\zeta(0)= -\frac{1}{2}$.
Finally, we define
\beq
	\left(\sum_{m\in\mathbb{Z}} \text{sgn}(\lam^{(\pm)}_m) \right)_{reg} = \eta_{\pm}(0)\ ,
\eeq
where $\eta_{\pm}(0)$ is the analytic continuation to $s=0$ of the eta function $\eta_{\pm}(s)$
of the operator $\mathcal{D}_{\pm}$ (see Appendix~\ref{app:dets} for details). We calculate
$\eta_{\pm}(0)$ in Appendix~\ref{app:dets} and we find that
\beq
	\eta_{\pm}(0)= \pm 1 \mp \frac{\overline{a}_t T}{\pi}\ .
\eeq

Putting this all together, we find that the regularized determinants of $\mathcal{D}_{\pm}$ are given by
\begin{align}
	\text{det}[\mathcal{D}_{\pm}]_{reg} &= 2\sin\left( \frac{\overline{a}_t T}{2}\right)e^{- i\frac{(2p+1)\pi}{2}(\pm 1\mp\frac{\overline{a}_t T}{\pi})} \nnb \\
	&= 2 (\mp i)^{2p+1} \sin\left( \frac{\overline{a}_t T}{2}\right) e^{\pm i\frac{(2p+1)}{2}\overline{a}_t T}\ ,
\end{align}
where $p$ was the arbitrary integer which appeared when we tried to rewrite $\text{sgn}(\lam^{(\pm)}_m)$ as an
exponential. We then find that the partition function for our quantum mechanical system coupled to the external field 
$A=A_t dt$ evaluates to
\beq
	Z[A]= (-1)^{k\ell + p+1} \frac{\sin\left[\frac{\overline{a}_t T}{2}(k-2p-1) \right]}{\sin\left(\frac{\overline{a}_t T}{2}\right)}\ .
\eeq

The next step is to determine which choice of $p$ gives the correct partition function. To do this, 
we will impose the following two
conditions on the value of $Z[A=0]$ (the partition function in zero external field). Physically, the value
of $Z[A=0]$ is the dimension of the Hilbert space of our quantum mechanical system. Therefore it makes sense to impose
the following two conditions on $Z[A=0]$.
\begin{enumerate}
\item For $k=0$, we require $Z[A=0]=1$, since $k=0$ gives a trivial theory with action equal to zero. The dimension of the 
Hilbert space of this theory should be equal to one.
\item For $k\neq 0$, $Z[A=0]$ should be a positive number.
\end{enumerate}
In terms of $\ell$ and $\overline{a}_t$, the limit $\overline{A}_t \to 0$ is taken by first setting $\ell=0$, and then taking
$\overline{a}_t \to 0$. In this limit we find
\beq
	Z[A=0]= (-1)^{p+1} (k - 2p -1)\ .
\eeq
The first condition implies that $p$ satisfies the equation
\beq
	1= (-1)^p(2p+1)\ .
\eeq
This equation has the two solutions $p=0$ and $p=-1$. For these two solutions for $p$, we find that $Z[A=0]$ at any
$k$ takes the form
\beq
	Z[A=0] = \begin{cases}
	-k+1 & ,\ p=0 \\
	k+1 & ,\ p=-1\\ 
\end{cases}\ .
\eeq
We see that in order to satisfy condition two, we must pick $p=-1$ for $k>0$ and $p=0$ for $k<0$. In this way we find that 
for all $k$, the partition function is given by
\beq
	Z[A]= (-1)^{k\ell } \frac{\sin\left[\frac{\overline{a}_t T}{2}(|k|+1) \right]}{\sin\left(\frac{\overline{a}_t T}{2}\right)}\ ,
\eeq
which is identical to the answer we computed for the $O(3)$ NLSM. Therefore we find that for \emph{any} two-dimensional 
target space $\M$ which respects the symmetries of the BTI phase, the NLSM description of the BTI using the target space 
$\M$ has \emph{the same} global anomaly as the $O(3)$ NLSM description. This result also implies that a large class
of bosonic theories in $0+1$ dimensions with $U(1)\rtimes\mathbb{Z}_2$  symmetry share the same global anomaly as a Dirac
fermion in $0+1$ dimensions with the same symmetry~\cite{elitzur1986origins}.

\section{Renormalization group flows and the fate of our models at low energies}
\label{sec:RG-flows}

In this section we briefly comment on the expected low energy behavior of the boundary theories discussed
in this article. Recall that the basic models we consider are NLSMs with a WZ term. On a $d$-dimensional spacetime $X_{bdy}$
(which we imagine to lie at the boundary of an SPT phase), we can
construct a WZ term for a NLSM with target space $\M$ if $\text{dim}[\M]= d+1$. In addition to the WZ term, the NLSM
action will also contain an ordinary kinetic term
\beq
	S_{kin}[\mb{m}]= \frac{1}{2f}\int_{X_{bdy}} d^d x\ G_{ab}(\mb{m})\pd_{\mu}m^a \pd^{\mu}m^b\ ,
\eeq
where $\mb{m}: X_{bdy}\to \M$ is the NLSM field, and $G_{ab}(\mb{m})$ is the Riemannian metric on $\M$ 
(compare with Eq.~\eqref{eq:kinetic-term-O(N)} for the case of a spherical target space). If we assume that
the NLSM field $\mb{m}$ is dimensionless, then the coupling constant $f$ has dimensions of $(\text{mass})^{2-d}$. 
Equivalently, the inverse $\frac{1}{f}$ of the coupling constant has dimensions of $(\text{mass})^{d-2}$.
We now consider the consequences of this fact for the low energy behavior of the theories discussed in this paper.
We focus on the case where $d\geq 2$ since for $d=1$ our theory is not a quantum field theory but just an ordinary quantum
mechanical system.

For simplicity, we first consider the case where the target space $\M$ is the sphere $S^{d+1}$ and so the 
NLSM field is a $(d+2)$-component unit vector $\mb{n}$. In the absence of
the WZ term (i.e., for a WZ term with level $k=0$) then for $d=2$ the renormalization group (RG) flow is towards
the disordered ($f\to\infty$) phase at all 
scales~\cite{polyakov1975interaction,brezin1976renormalization,brezin1976spontaneous,hamer1979strong}. In this
limit the theory is massive and the ground state (or vacuum state) possesses the full $O(d+2)=O(4)$ symmetry of the action 
(the ground state transforms as a singlet under the action of the $O(4)$ symmetry). When the WZ
term is turned on, a stable fixed point appears at a finite value of the coupling $f$~\cite{witten1984non},
and this fixed point is actually the $SU(2)_k$ Wess-Zumino-Witten conformal field theory. To see this we note that the 
four-component unit vector field of the $O(4)$ NLSM is equivalent to a $2\times 2$ $SU(2)$ matrix field. Explicitly, if 
$\mb{n}= (n^1,\dots,n^4)$, then one possible mapping to the matrix field $U$ is
$U= n^4\mathbb{I} + i \sum_{a=1}^3 n^a \sigma^a$, where $\sigma^a$  for $a=1,2,3,$ are the Pauli matrices.
In addition, the $U(1)$ symmetry that we are interested in in this paper is realized as a right (or left, depending on the
mapping from $\mb{n}$ to $U$) $U(1)$ symmetry of the $SU(2)_k$ theory, and this symmetry is well-known to be
anomalous~\cite{WittenHolo,HullSpence1}.

For the case of $d>2$ the coupling constant $f$ is dimensionful and one expects (by a simple power-counting argument)
that the theory flows towards the ordered phase $f\to 0$ and so the $O(d+2)$ symmetry of the theory is spontaneously
broken at low energies. In fact, for the theory without a topological term, a double perturbation expansion in $f$ and 
$\ep= d-2$ reveals the existence of an unstable fixed point at a finite value $f_1$ of the (suitably rescaled) coupling 
$f$~\cite{brezin1976renormalization,brezin1976spontaneous}. If this computation can be trusted, then below this fixed
point the theory flows to $f\to 0$ and the symmetry is spontaneously broken, while above the fixed point the theory flows
to a (presumably) disordered strong-coupling ($f\to\infty$) phase in which the $O(d+2)$ symmetry is 
restored\footnote{The present authors provided further evidence for the existence of this strong coupling phase in our recent 
work~\cite{lapa2017canonical}, where we computed the beta function for the coupling constant $f$ to leading order in a
strong-coupling lattice regularization inspired by the approach of Ref.~\onlinecite{hamer1979strong}.}. Since in the 
$d=2$ case turning on the WZ term introduces a stable fixed point at a finite value of the coupling, some authors
have recently proposed a scenario for $d>2$ in which the introduction of the WZ term introduces a stable fixed point at a finite 
value $f=f_2$ of the coupling constant, with $f_2>f_1$, where $f_1$ is the location of the unstable fixed 
point (see Figure 2a of Ref.~\onlinecite{xu2013nonperturbative}). 
This possibility was first raised in Ref.~\onlinecite{xu2013nonperturbative}, and it has been pursued recently in 
Ref.~\onlinecite{bi2016stable} using a combination of several perturbation expansions. Both of these works consider the
case of $d=3$ spacetime dimensions\footnote{More precisely, Ref.~\onlinecite{xu2013nonperturbative} considered
the $O(4)$ NLSM in $d=3$ spacetime dimensions with a topological \emph{theta term} with coefficient $\pi$. This theory can be 
understood as a deformation of the $O(5)$ NLSM with WZ term (in the same dimension) at level $k=1$ in which the fifth
component of the NLSM field has been set to zero.}. 

What can we deduce about our boundary theories from this discussion? Let us first consider the
case for BIQH states. Recall that these boundary theories were $O(2m)$ NLSMs with
WZ term in spacetime dimension $2m-2$, and also NLSMs with deformed target spaces $\M$ which still possessed a 
$U(1)$ symmetry. We first discuss the case $m>2$, so that the boundary spacetime dimension is larger than two. 
In this case, the conclusion which is supported by the most evidence is that the $U(1)$ symmetry of these theories is 
spontaneously broken in the ground state. In this case the symmetry-broken theory will possess a gapless Goldstone mode. 
Interestingly, this gapless mode will still couple to the external field $A=A_{\mu}dx^{\mu}$ and it is this Goldstone
mode which exhibits the anomaly in the symmetry-broken theory. For example,
if we consider a general target space $\M$ with $U(1)$ symmetry, and we add a potential to the action which is minimized
along the $U(1)$ orbit of a particular point on $\M$, then the low energy theory will possess a gapless Goldstone mode
corresponding to motion around this orbit\footnote{We would like to thank one of the referees of this paper for suggesting 
that we consider an example of this kind.}. 

It is helpful to see an explicit example of this kind in order to appreciate the fact that the Goldstone mode really does 
exhibit the anomaly.  Let us take the $O(2m)$ NLSM with WZ term at
level $k$ and introduce a potential into the action which is minimized when $|b_1|^2= 1$ and all other $b_{\ell}=0$
($\ell=2,\dots,m$). In the symmetry-broken vacuum we then have $b_1= e^{i\vphi_{vac}}$ for some constant
$\vphi_{vac}$. If we expand around this vacuum by setting $b_1= e^{i\vphi_{vac}+i\vphi}$ then the gauged NLSM 
action with WZ term reduces to an action for the gapless Goldstone mode $\vphi$ coupled to $A$. This action takes the explicit 
form
\beqa
	S[\vphi] &=& \frac{1}{2f}\int d^{2m-2}x\ (\pd_{\mu}\vphi- A_{\mu})(\pd^{\mu}\vphi- A^{\mu}) \nnb \\ 
	&-& \frac{k}{(2\pi)^{m-1}}\int_{X_{bdy}} d\vphi \wedge A\wedge F^{m-2}\ .
\eeqa
It is now easy to see that under a $U(1)$ gauge transformation $\vphi \to \vphi + \chi$, $A\to A+ d\chi$, the action for
the Goldstone mode $\vphi$ has the same anomaly as the original $O(2m)$ NLSM. From this analysis we can conclude that
even when the $U(1)$ symmetry of the BIQH state is spontaneously broken in the boundary theory, the boundary theory
will still possess the same perturbative $U(1)$ anomaly as the original NLSM that we started with.

In the case of $m=2$ (boundary spacetime dimension equal to two) the situation is more interesting. As we noted above,
if we preserve the full $O(4)$ symmetry of the theory, then our theory flows at low energies to the $SU(2)_k$
Wess-Zumino-Witten conformal field theory. On the other hand, we can introduce some $O(4)$-breaking but 
$U(1)$-preserving anisotropy into the theory to set $|b_1|^2= 1$ and $b_2=0$ (or vice-versa). 
In this case we end up with a free boson theory of the form
\beqa
	S[\vphi] &=& \frac{1}{2f}\int d^{2}x\ (\pd_{\mu}\vphi- A_{\mu})(\pd^{\mu}\vphi- A^{\mu}) \nnb \\ 
	&-& \frac{k}{2\pi}\int_{X_{bdy}}d\vphi\wedge A\ ,
\eeqa
where we have $b_1=e^{i\vphi}$. 
In this case, however, $\vphi$ should not be interpreted as a Goldstone boson as we do not have spontaneous symmetry
breaking in this dimension. The $SU(2)_k$ theory has a central charge of $c= \frac{3k}{k+2} \geq 1$ 
(see, for example, Ref.~\onlinecite{CFTbook}) so it can and will flow to 
the free boson theory with central charge $c=1$ when perturbations which break the $O(4)$ symmetry down to $U(1)$ are 
introduced (this flow is consistent with Zamolodchikov's c-theorem~\cite{zamolodchikov1986}). Note that
if we preserve the $U(1)$ symmetry, then the boundary theory cannot be gapped out since we always need some gapless
degrees of freedom to saturate the anomaly. Finally, we remark that in the $k=1$ case, the free boson theory is actually
equivalant to the $SU(2)_1$ theory for a particular value of the coupling $f$. However, marginal perturbations which break
the $O(4)$ symmetry down to $U(1)$ will in general tune $f$ away from this special value.

We close this section with a few words about the boundary theories of BTI states. These boundary theories occur in
odd spacetime dimensions $2m-1$, and they lie at the boundary of a BTI state in $2m$ dimensions. We have already
analyzed the case $m=1$ in detail in Sec.~\ref{sec:global}. In this case the boundary is just a quantum mechanical system
and there are no subtleties involved in assessing the fate of the system at low energies. For the case of $m>1$, the
most likely scenario is that these boundary theories spontaneously break the $U(1)\rtimes\mathbb{Z}_2$ symmetry of the
BTI state. As we noted in Ref.~\onlinecite{lapa2017}, because of the way the $U(1)$ symmetry in our models acts on 
$S^{2m}$ (the target space of the NLSM in this case), in the BTI case it is possible to break the $\mathbb{Z}_2$ symmetry 
while preserving the $U(1)$ symmetry. In this 
way we were able to show that the boundary of the BTI state can exhibit a $\mathbb{Z}_2$ symmetry-breaking 
electromagnetic response, and we found that this response is given by a CS term for $A_{\mu}$ with level $\frac{m!}{2}$. 
We then argued, based on this evidence, that the boundary theories of the BTI state exhibit a bosonic analogue of the
parity anomaly. 

For $m>1$ it is still an open problem to exhibit this bosonic analogue of the parity anomaly in a concrete way (e.g., at the level 
of the partition function). The most interesting case is $m=2$ in which the boundary spacetime dimension is $d=3$. 
Here we can list three possibilities for the fate of the boundary theory at low energies. First, as noted above, the boundary 
could break part or all of the symmetry group $U(1)\rtimes\mathbb{Z}_2$ of the BTI state. Second,  
the results of Refs.~\onlinecite{xu2013nonperturbative,bi2016stable} indicate that
a gapless conformal field theory preserving the full $U(1)\rtimes\mathbb{Z}_2$ symmetry may be possible. Finally, since the
anomaly in this case is global and not perturbative, there is the possibility that the boundary theory can flow to a 
topological quantum field theory whose partition function (in the presence of the external field $A_{\mu}$) exhibits the
anomaly. In this last case all other degrees of freedom at the boundary become gapped and decouple from the topological
quantum field theory which describes only the ground state sector of the boundary theory. We comment more on this last 
possibility in Sec.~\ref{sec:conclusion}.

\section{Discussion and Conclusion}
\label{sec:conclusion}

In this paper we continued the program, initiated in Ref.~\onlinecite{lapa2017}, of characterizing the anomalies at the
boundary of BIQH and BTI states in all odd and even dimensions, respectively. In Sec.~\ref{sec:pert} we 
revisited the perturbative $U(1)$ anomaly at the boundary of BIQH states. There we proved that the
target space $\M$ of the NLSM describing the boundary theory of these states can be subjected to arbitrary smooth, 
symmetry-preserving deformations without affecting the anomaly. 
In Sec.~\ref{sec:global} we revisited the global anomaly at the boundary of BTI states. In Ref.~\onlinecite{lapa2017} we 
gave an argument that the boundary of the BTI state exhibits a bosonic analogue of the parity anomaly of Dirac fermions in 
odd dimensions. In this paper we elevated this argument to a proof for the case of the $(0+1)$-dimensional boundary of the 
$(1+1)$-dimensional BTI state. In that case we also used the equivariant localization technique to prove that the global 
anomaly of the BTI boundary is robust against arbitrary smooth, symmetry-preserving deformations of the target space of the 
NLSM used to describe this state.

From a fundamental point of view, perhaps the most important result in this paper is our concrete demonstration, at the 
level of the partition function, of an analogue of the parity anomaly in a purely bosonic system. Indeed, our result in 
Sec.~\ref{sec:global} is a direct bosonic analogue of the results of Ref.~\onlinecite{elitzur1986origins} on global anomalies of 
fermions in $0+1$ dimensions. In the context of SPT phases, our results in this paper also imply that the universal properties of 
an SPT phase can be captured by a much wider range of models than the NLSMs with spherical target space originally 
considered in Refs.~\onlinecite{CenkeClass1,CenkeClass2}. 
The results of this paper lead us to conjecture that an SPT phase in $D+1$
dimensions with symmetry group $G$, which would be described by an $O(D+2)$ NLSM in the approach of 
Refs.~\onlinecite{CenkeClass1,CenkeClass2}, can be modeled using an NLSM with \emph{any} target space $\M$
related to $S^{D+1}$ by a diffeomorphism which is equivariant with respect to the action of the group $G$. Note that this
conjecture only applies to SPT phases for which an NLSM description \emph{exists}. This does not seem to be the
case for all SPT (or short-range entangled) phases, for example the ``E8" state in $2+1$ dimensions and the ``beyond 
cohomology" state with time-reversal symmetry in $3+1$ dimensions~\cite{VL2012,VS2013}. 

An ambitious goal for future work would be to present a concrete demonstration, again at the level of the partition function,
of an analogue of the parity anomaly in a $(2+1)$-dimensional bosonic model with $U(1)$ and $\mathbb{Z}_2$ symmetry,
where $\mathbb{Z}_2$ now represents time-reversal. A precise understanding of global anomalies in $(2+1)$-dimensional bosonic 
systems would also be extremely useful in the search for new dualities in quantum field theory in $2+1$ 
dimensions~\cite{son2015composite,wang2015dual,metlitski2016particle,mross2016explicit,karch2016particle,seiberg2016duality,wang2017deconfined}. 
A crucial check on any proposed duality is that the two theories which are conjectured to be dual to each other must have the
same `t Hooft anomalies when coupled to various external fields.  

An interesting candidate for a $(2+1)$-dimensional bosonic model displaying a bosonic analogue of the parity anomaly is the 
$O(5)$ NLSM with WZ term, and with the $U(1)\rtimes\mathbb{Z}_2$ symmetry of the BTI state acting in the manner described 
in Ref.~\onlinecite{lapa2017}. In Ref.~\onlinecite{lapa2017} we already gave several pieces of evidence which suggest that 
this model displays a bosonic analogue of the parity anomaly. The first piece of evidence was our computation of the 
time-reversal breaking electromagnetic response of this model, which we already mentioned above. However, we also gave a 
second argument which was based on the demonstration that there is a certain composite vortex excitation in this model with 
fermionic statistics (an observation which goes back to Refs.~\onlinecite{SenthilFisher,VS2013}), and such an excitation should 
not exist in a purely bosonic model which is not anomalous. 

The $O(5)$ NLSM with WZ term may be tractable analytically in the topological limit in which the 
coupling constant $f_{bdy}$ of the NLSM is sent to infinity (i.e., if one considers the model with only the topological term). 
This would correspond to the third possibility that we raised at the end of Sec.~\ref{sec:RG-flows}: the 
boundary theory could flow to a topological quantum field theory whose partition function exhibits the anomaly.
It may even be the case that a more sophisticated version of the equivariant localization 
technique can be used to calculate the partition function of the $O(5)$ NLSM with WZ term in the topological limit and properly 
coupled to an external $U(1)$ gauge field as described in Sec.~VI of Ref.~\onlinecite{lapa2017}. However, there are 
several difficulties which must be surmounted before one can apply any kind of equivariant localization technique to this
problem. The main problem is that one needs to find a hidden supersymmetry in this problem which can be exploited in order
to establish the localization of the path integral. In the $(0+1)$-dimensional case this supersymmetry followed, at
least partially\footnote{As we reviewed in Appendix~\ref{app:EL}, the fact that the Hamiltonian was associated with a
$U(1)$-action on the phase space was also a crucial ingredient.}, from the fact that the path integral measure could be 
exponentiated by introducing a set of \emph{real} Grassmann-valued (i.e., fermionic) fields $\eta^a(t)$ into the problem. 
This could only be done with real fermionic fields because the target spaces of the $(0+1)$-dimensional NLSMs that we 
studied were all symplectic manifolds. On the other hand, the target space $S^4$ of the $O(5)$ NLSM is not symplectic.
Therefore one can only exponentiate the path integral measure by introducing complex fermionic fields. Currently, we are not
aware of a generalization of the equivariant localization techniques of 
Refs.~\onlinecite{blau1990path,keski1991topological,niemi1992cohomological,niemi1994exact} which starts by exponentiating
the path integral measure by introducing complex fermions, but such a generalization may still be possible.  
We leave a detailed investigation of this to future work.

\acknowledgements

We thank M. Stone and T. Zhou for helpful discussions on the content of this paper. MFL would also like to acknowledge the 
hospitality of the Galileo Galilei Institute for Theoretical Physics in Florence, Italy, where part of this work was 
completed. MFL and TLH acknowledge support from the ONR YIP Award N00014-15-1-2383.  We also gratefully acknowledge 
the support of the Institute for Condensed Matter Theory at the University of Illinois at Urbana-Champaign.

\appendix

\section{Classical mechanics and phase space path integral for general Hamiltonian systems}
\label{app:symp-geom}

In this appendix we review the symplectic geometry formulation of classical Hamiltonian mechanics, closely following the
discussion in Ch.~11 of Ref.~\onlinecite{stonebook}. We use this formalism in Sec.~\ref{sec:global} of the paper to aid in the 
evaluation of the partition function for a gauged NLSM with WZ term which describes the $(0+1)$-dimensional boundary of the BTI 
state in $1+1$ dimensions. The symplectic geometry formulation of Hamiltonian mechanics is a geometric formulation in terms of a 
phase space $\mathcal{M}$ (a closed, orientable, smooth manifold) equipped with a symplectic form $\omega$. We take 
$\mathcal{M}$ to have dimension $2n$, where $n$ is an integer greater than or equal to one. The symplectic form
$\omega$ is a closed, non-degenerate two-form on $\mathcal{M}$. In a system of local coordinates $m^a$ on $\mathcal{M}$, in 
which $\omega= \frac{1}{2}\omega_{ab}(\mb{m}) dm^a\wedge dm^b$, the non-degeneracy condition is equivalent to the 
condition that the components $\omega_{ab}(\mb{m})$ are the elements of an invertible matrix. 
We use the notation $\mb{m}= (m^1,\dots,m^{2n})$ to refer to the entire collection of phase space coordinates, and we use 
Latin indices near the beginning of the alphabet to label the components of general tensor fields on $\mathcal{M}$. We also use 
the notation $\pd_a \equiv \frac{\pd}{\pd m^a}$ and $\dot{m}^a \equiv \frac{d m^a}{dt}$ in what follows.

To start, for any function $f$ on phase space we define an associated vector field $\un{v}_f$ by the 
equation
\beq
	df= -i_{\un{v}_f}\omega\ , \label{eq:function-vector}
\eeq
where $i_{\un{v}}\omega = v^{a}\omega_{ab}dm^b$ denotes interior multiplication of the form $\omega$ by
the vector field $\un{v}$. The components of $\un{v}_f$ then take the form
\beq
	v^{a}_f= \omega^{ab}\pd_{b}f\ ,
\eeq
where $\omega^{ab}$ are the elements of a matrix which is the inverse of the matrix with elements $\omega_{ab}$,
i.e.,
\beq
	\omega^{ab}\omega_{bc}= {\delta^{a}}_{c}\ .
\eeq
We see that the symplectic two-form $\omega$ must be non-degenerate for this to work.

The Poisson bracket of two functions $f$ and $g$ on phase space is then defined by\footnote{The placement of $\un{v}_f$ and
$\un{v}_g$ on the right-hand side of this equation is not a typo. We are using the non-traditional definition of the Poisson bracket from
Ref.~\onlinecite{stonebook}.}
\beq
	\{f,g\} = i_{\un{v}_g}i_{\un{v}_f}\omega\ .  \label{eq:Poisson-bracket} 
\eeq
In a system of local coordinates the Poisson bracket has the form
\beq
	\{f,g\}  = \omega^{ab}\pd_{b}f\pd_{a}g\ .
\eeq
For a given Hamiltonian function $H$, Hamilton's equations are equivalent to the single equation
\beq
	dH= -i_{\un{v}_H}\omega\ ,
\eeq
where $\un{v}_H$ is the vector field whose components are the time derivatives of the phase space coordinates,
\beq
	\un{v}_H= \dot{m}^{a}\pd_{a}\ .
\eeq
Finally, in each coordinate patch on $\mathcal{M}$ we can write 
\beq
	\omega=d\vth \ , \label{eq:symp-potential}
\eeq
where the one-form $\vth= \vth_a(\mb{m})dm^a$ is known as the \emph{symplectic potential}. 

Next, we review the form of the phase space path integral for the partition function $Z(T)= \text{tr}[e^{-iHT}]$ of the
quantum mechanical system obtained via quantization of the classical system defined by the triple 
$(\mathcal{M},\omega,H)$. Here the trace is over the Hilbert space of the quantum
mechanical system. As is reviewed in Sec.~4.1 of Ref.~\onlinecite{szabo2003equivariant}, the phase space
path integral for $Z(T)$ takes the form
\beq
	Z(T) = \int_{L\mathcal{M}} [d^{2n}\mb{m}] \left[\prod_{t\in[0,T)} \text{Pf}[\omega_{ab}(\mb{m}(t))]\right] e^{iS[\mb{m}]}\ , \label{eq:partition-fcn}
\eeq
where the action appearing in the exponential is
\beq
	S[\mb{m}]= \int_0^T dt\ \left[ \vth_{a}(\mb{m})\dot{m}^{a} - H(\mb{m}) \right]\ .
\eeq
The path integral is taken over all field configurations $m^a(t)$ with periodic boundary conditions $m^a(0)= m^a(T)$ on the
interval $[0,T)$. The space of all such configurations is known as the \emph{loop space} $L\mathcal{M}$ of the 
phase space manifold $\mathcal{M}$. In addition, $[d^{2n}\mb{m}]$ denotes a flat measure on phase space at all points in time. 
The nontrivial geometry of the phase space is taken into account by the insertion of
\beq
	\prod_{t\in[0,T)} \text{Pf}[\omega_{ab}(\mb{m}(t))]
\eeq
into the path integral. This factor can be understood by noting that the $2n$-form $\frac{\omega^n}{n!}$ provides a natural 
volume form (the Liouville measure) on $\mathcal{M}$, and also by making use of the formula 
$\frac{\omega^n}{n!}= \text{Pf}[\omega_{ab}] dm^1\wedge\cdots\wedge dm^{2n}$.

The first term in the action can also be recast into a form which is very similar to a WZ term. 
Let us denote the interval $[0,T)$ with periodic boundary conditions by $S^1_T$, the circle of circumference $T$. This circle
is the spacetime that our quantum mechanical system evolves on. To write the first term in the action in a WZ form, we
first introduce a two-dimensional manifold $\B$ which has $S^1_T$ as its boundary, $\pd\B= S^1_T$. Then we choose an extension 
$\tilde{\mb{m}}$ of the field configuration $\mb{m}$ into the bulk of $\B$ such that $\tilde{\mb{m}}|_{\pd\B}= \mb{m}$. We can
now use Stokes' theorem to rewrite the first term in $S[\mb{m}]$ as
\beqa
	\int_0^T dt\ \vth_{a}(\mb{m})\dot{m}^{a} &=& \int_{S^1_T}\mb{m}^*\vth \nnb \\
	&=& \int_{\B}\tilde{\mb{m}}^*d\vth \nnb \\
	&=& \int_{\B}\tilde{\mb{m}}^*\omega\ .
\eeqa
In this form the term $\int_0^T dt\ \vth_{a}(\mb{m})\dot{m}^{a}$ appearing in the action looks very similar to a WZ
term, in the sense that it involves (i) an extended spacetime $\B$, (ii) an extension $\tilde{\mb{m}}$ of the field configuration 
$\mb{m}$ into $\B$, and (iii) the integral over $\B$ of the pullback of a \emph{closed} form on $\M$.

\section{A brief introduction to equivariant localization for phase space path integrals}
\label{app:EL}

In this appendix we give a brief review of the \emph{equivariant localization} (EL) technique for the evaluation of certain phase 
space path integrals of the form of Eq.~\eqref{eq:partition-fcn} from Appendix~\ref{app:symp-geom}. 
We use the EL technique in Sec.~\ref{sec:global} to evaluate the partition function for a gauged NLSM with WZ term which 
describes the $(0+1)$-dimensional boundary of a BTI state in $1+1$ dimensions. Our presentation in this appendix is based on
the discussion in Sec.~4 of Ref.~\onlinecite{szabo2003equivariant}. We also give a  brief discussion on how one can
define the Pfaffians of infinite-dimensional operators which appear in the formulas obtained by applying the EL technique.

The EL technique for phase space path integrals can be thought of as an infinite-dimensional generalization of the finite-dimensional
integration formulas derived in Refs.~\onlinecite{duistermaat1982variation,berline1982classes,atiyah1984moment}. 
In this paper we only use the simplest version of the EL technique. The path integral formula which follows from this particular
version of the EL technique is sometimes referred to as the ``WKB" localization formula. 
This basic version of the EL method and several generalizations of it (in particular the ``Niemi-Tirkkonen" formula) were developed 
in Refs.~\onlinecite{blau1990path,keski1991topological,niemi1992cohomological,niemi1994exact}. Stone's
paper~\cite{stone1989supersymmetry} on a hidden supersymmetry in the quantum mechanics of spin can be seen as a herald
for the developments on the EL technique for phase space path integrals which followed soon after. The application of the EL 
technique to systems with a two-dimensional phase space, which is the case of interest in this paper, was considered in detail in 
Ref.~\onlinecite{szabo1994phase}. Finally, some issues related to the regularization of determinants and Pfaffians 
appearing in the EL formulas were greatly clarified by Miettinen in Ref.~\onlinecite{miettinen1996localization}. 

In the context of the EL technique, the word ``localization" refers to the fact that although the path integral in question 
ostensibly gets contributions from all possible field configurations, the final
result only depends on contributions from a very small subset of these configurations. Thus, the integral ``localizes" 
to a sum or, in some cases, a finite-dimensional integral over this subset of all field configurations. The word 
``equivariant" refers to the fact that the mechanism responsible for the localization of the integral is best understood in
terms of the \emph{equivariant cohomology} of the manifold that one is integrating over~\cite{atiyah1984moment}. In the case of 
the phase space path integrals considered here this turns out to be the $U(1)$-equivariant cohomology of the infinite-dimensional 
\emph{loop space} $L\M$ of the classical phase space $\M$.

The basic idea of the EL technique is as follows. 
First, to apply the EL technique we need to start with a classical system possessing a $U(1)$ 
symmetry. It turns out that this $U(1)$ symmetry ``lifts" to 
a supersymmetry of the phase space path integral. This supersymmetry is then used to construct a one parameter family of
equivalent path integrals parametrized by $\lambda \in [0,\infty)$, with the original path integral of interest corresponding to 
$\lambda=0$. The supersymmetry guarantees that the path integral at any value of $\lam$ is equivalent to the original 
path integral. Therefore, the original path integral can be computed by taking the opposite limit $\lam\to\infty$.
In this limit the path integral simplifies dramatically, getting contributions only 
(in the cases considered here) from the field configurations which correspond to \emph{time-independent} 
solutions to the classical equations of motion. One says that the path integral \emph{localizes} onto these configurations.
We now outline the main ideas behind the EL technique in more detail, closely following Sec.~4 of 
Ref.~\onlinecite{szabo2003equivariant}. 

To start, we assume that it is possible to define an action of the group $U(1)$ on the phase space $\mathcal{M}$. 
Let $\un{v}= v^a(\mb{m})\pd_a$ be the vector field which generates the $U(1)$ action, in the sense that under
a $U(1)$ transformation by the small angle $\xi$ the phase space coordinates transform as $m^a \to m^a + \xi v^a$. 
On $\mathcal{M}$ there is a Hamiltonian function $H(\mb{m})$ which 
is naturally associated with this vector field, and which is determined by $\un{v}$ and $\omega$ via the equation
\beq
	dH= -i_{\un{v}}\omega\ .
\eeq
Note that this is just Eq.~\eqref{eq:function-vector} with the function $f$ taken to be the Hamiltonian.
We choose this specific Hamiltonian to describe the dynamics of the system that we consider in 
what follows. With this choice of Hamiltonian, the action for our dynamical system will also have a $U(1)$ symmetry.
Finally, we will need a Riemannian metric $g_{ab}(\mb{m})$ on $\M$ which is invariant under the $U(1)$ action generated by 
$\un{v}$. This is equivalent to the requirement that $\un{v}$ is a Killing vector for the metric, i.e., $g_{ab}$ and $v^a$ should 
satisfy the Killing equation
\beq
	v^c \pd_c g_{ab} + g_{ac}\pd_b v^c + g_{bc}\pd_a v^c = 0, \forall\ a,b\ .
\eeq

The path integral in Eq.~\eqref{eq:partition-fcn} involves an integration over the loop space $L\mathcal{M}$ of $\mathcal{M}$,
which is spanned by the $T$-periodic functions $m^a(t)$ which, for each $t$, represent a point on $\mathcal{M}$.
We now introduce an additional set of Grassmann-valued fields $\eta^a(t)$ which also obey periodic boundary conditions. The 
space of these new fields is equivalent to the loop space of $\Lambda^1\mathcal{M}$, the vector space of one-forms on 
$\mathcal{M}$, and this space is denoted by 
$L\Lambda^1\mathcal{M}$. The interpretation in terms of $\Lambda^1\mathcal{M}$ is due to the fact that at each time
$t$ the anticommuting fields $\eta^a(t)$ can be regarded as a basis of one-forms on $\mathcal{M}$.
Using the rules for integration over real Grassmann variables, the new fields $\eta^a(t)$ can
be used to rewrite $Z(T)$ in the form 
\beq
	Z(T)= \int_{L\mathcal{M}\otimes L\Lambda^1\mathcal{M}} [d^{2n}\mb{m}] [d^{2n}\mbs{\eta}]\ e^{i(S[\mb{m}] + \Omega[\mb{m},\mbs{\eta}])} \ , \label{eq:partition-with-grassmann}
\eeq
where we defined 
\beq
	\Omega[\mb{m},\mbs{\eta}]= \frac{1}{2}\int_0^T dt\ \omega_{ab}(\mb{m}(t))\eta^a(t)\eta^b(t)\ ,
\eeq
and where the integration is now over the ``super loop space" $L\mathcal{M}\otimes L\Lambda^1\mathcal{M}$.
One should compare Eq.~\eqref{eq:partition-with-grassmann} with the original expression 
Eq.~\eqref{eq:partition-fcn} for $Z(T)$.

Using the Grassmann-valued fields we can define the operators
\beq
	d_L= \int_0^T dt\ \eta^a(t)\frac{\delta}{\delta m^a(t)}\ ,
\eeq
and 
\beq
	i_{S}= \int_0^T dt\ V^a_S[\mb{m}(t);t] \frac{\delta}{\delta\eta^a(t)}\ ,
\eeq
where
\beq
	V^a_S[\mb{m}(t);t]= \dot{m}^a(t) - v^a(\mb{m}(t))\ .
\eeq
The quantities $V^a_S[\mb{m}(t);t]$ can be interpreted as the components of a vector field 
\beq
	\un{V}_S= \int_0^T dt\ V^a_S[\mb{m}(t);t] \frac{\delta}{\delta m^a(t)}
\eeq
on the loop space. To understand the physical significance of the components $V^a_S[\mb{m}(t);t]$, 
note that the classical equations of motion for the system under consideration are
\beq
	\frac{\delta S[\mb{m}]}{\delta m^a(t)}= \omega_{ab}(\mb{m}(t))V^b_S[\mb{m}(t);t] = 0\ ,\ \forall\ a\ . \label{eq:classical-EOMs}
\eeq
Since $\omega$ is non-degenerate, the classical equations of motion are equivalent
to the equations $V^a_S[\mb{m}(t);t]= 0,\ \forall\ a$. 
The operator $d_L$ can be interpreted as an exterior derivative on $L\mathcal{M}$,
and $i_S$ has the interpretation of interior multiplication by the loop space vector field $\un{V}_S$. 

In terms of these operators we now define the loop space equivariant exterior derivative 
\beq
	Q_S= d_L + i_S\ .
\eeq
The square of this operator can be interpreted as a loop space Lie derivative (acting on loop space differential forms) 
along the loop space vector field $\un{V}_S$,
\beq
	\mathcal{L}_S \equiv Q^2_S= d_L i_S + i_S d_L\ .
\eeq
Some algebra shows that 
\beq
	Q_S( S[\mb{m}] + \Omega[\mb{m},\mbs{\eta}]) =  0\ ,\label{eq:closure}
\eeq
which means that the integrand in the path integral is equivariantly closed (i.e., closed under the action of the equivariant
exterior derivative). To prove this relation one needs to use the fact that $\omega$ is closed as an ordinary two-form on 
$\mathcal{M}$, and also Eq.~\eqref{eq:classical-EOMs} relating $\un{V}_S$ to the classical equations of motion.  

The closure of the integrand can be interpreted in terms of a supersymmetry (SUSY) of this system which is generated by the 
``supercharge" $Q_S$. In particular, Eq.~\eqref{eq:closure} implies that the path integral for $Z(T)$ is invariant under the 
SUSY transformation
\begin{subequations}
\label{eq:SUSY-trans}
\beqa
	\delta_{\ep} m^a(t) &=& \ep Q_S m^a(t) \\
	\delta_{\ep} \eta^a(t) &=& \ep  Q_S \eta^a(t)\ ,
\eeqa
\end{subequations}
where $\ep$ is a constant Grassmann parameter. An explicit calculation gives
$Q_S m^a(t)= \eta^a(t)$ and $Q_S \eta^a(t)= V^a_S[\mb{m}(t);t]$, so we know the exact form that this SUSY 
transformation takes. The next step towards establishing localization of the path integral
is to use the supersymmetry to deform the path integral by adding a suitably chosen SUSY-exact term to the integrand. 
To this end, we modify $Z(T)$ to
\begin{align}
	Z(T,&\lambda)  \nnb \\
 = &\int_{L\mathcal{M}\otimes L\Lambda^1\mathcal{M}} [d^{2n}\mb{m}] [d^{2n}\mbs{\eta}]e^{i(S[\mb{m}] + \Omega[\mb{m},\mbs{\eta}]) - \lambda Q_S\Psi[\mb{m},\mbs{\eta}]}\ ,
\end{align}
where $\Psi[\mb{m},\mbs{\eta}]$ is some functional of $\mb{m}$ and $\mbs{\eta}$ which will be required to 
satisfy 
\beq
	Q_S^2\Psi[\mb{m},\mbs{\eta}]= 0\ .
\eeq

If we can find such a functional $\Psi[\mb{m},\mbs{\eta}]$, then we can show that $Z(T,\lam)$ is actually independent of 
$\lam$ by the following manipulations. We compute (we suppress the arguments of the different terms for brevity)
\beqa
	\frac{d Z(T,\lambda)}{d\lam} &=&  -\int [d^{2n}\mb{m}] [d^{2n}\mbs{\eta}]\ Q_S\Psi\ e^{i(S + \Omega) - \lambda Q_S\Psi} \nnb \\
	&=& -\int [d^{2n}\mb{m}] [d^{2n}\mbs{\eta}]\ Q_S \left[ \Psi\ e^{i(S + \Omega) - \lambda Q_S\Psi} \right] \nnb \\
	&=& 0\ .
\eeqa
The second line follows from the first since the argument of the exponential is annihilated by $Q_S$ (and this requires
that $Q_S^2\Psi= 0$). Finally, the third line follows from the second due to an infinite-dimensional version of the statement
that the integral of a total derivative is zero. In the infinite-dimensional case this is only true if the path integral measure is
invariant under the action of $Q_S$, but that is the case here. An alternative explanation of the $\lam$-independence
of this integral, which uses a Ward identity associated with the symmetry generated by $Q_S$, can be found in 
Ref.~\onlinecite{szabo2003equivariant}.

The arguments from the last paragraph show that the original partition function $Z(T)$ is equal to the deformed partition
function $Z(T,\lam)$ for any value of $\lam$. The final step in establishing the localization of $Z(T)$ is to pick a particular
functional $\Psi[\mb{m},\mbs{\eta}]$ such that the $\lam\to\infty$ limit of $Z(T,\lam)$ becomes easy to evaluate.
There are various choices for such a $\Psi[\mb{m},\mbs{\eta}]$, but the choice which leads to the WKB localization formula
is 
\beq
	\Psi[\mb{m},\mbs{\eta}]= \int_0^T dt\ g_{ab}(\mb{m}(t))V^a_S[\mb{m}(t);t]\eta^b(t)\ .
\eeq
One can check that this functional satisfies $Q_S^2\Psi[\mb{m},\mbs{\eta}]= 0$, but the derivation relies on the 
fact that $\un{v}$ is a Killing vector for the metric $g_{ab}$. 

Using this particular choice of $\Psi[\mb{m},\mbs{\eta}]$, one can now show that 
the path integral $Z(T)$ localizes to a sum over contributions from the field configurations in the set
\beq
	L\mathcal{M}_S= \left\{ \mb{m}(t) \in L\mathcal{M}\ |\ V^a_S[\mb{m}(t);t]= 0,\ \forall\ a  \right\}\ , \label{eq:set-of-periodic-solutions}
\eeq
which is the set of all $T$-periodic solutions to the classical equations of motion. To motivate this, we simply note that
the bosonic term in $Q_S\Psi[\mb{m},\mbs{\eta}]$ is
\beq
	\int_0^T dt\ g_{ab}(\mb{m}(t)) V^a_S[\mb{m}(t);t] V^b_S[\mb{m}(t);t] \ .
\eeq
Now $Q_S\Psi[\mb{m},\mbs{\eta}]$ appears in the exponential of the path integral multiplied by a factor of $-\lam$, which
means that in the limit $\lam\to\infty$, this term becomes a delta function which restricts the path integral to 
only those field configurations where $V^a_S[\mb{m}(t);t]= 0$.

The final result of the EL calculation is the formula
\beqa
	Z(T) &=& \lim_{\lam\to\infty} Z(T,\lam) \nnb \\
	&\sim& \sum_{\overline{\mb{m}}(t) \in  L\mathcal{M}_S} \frac{e^{iS[\overline{\mb{m}}]}}{\text{Pf}[\mathcal{O}]_{\mb{m}(t)=\overline{\mb{m}}(t)}}\ , \label{eq:EL-result}
\eeqa
where the infinite-dimensional operator $\mathcal{O}$ has matrix elements
\begin{align}
	\mathcal{O}_{ab}(t,t') &= \frac{\delta V^c_S[\mb{m}(t');t']}{\delta \mb{m}^a(t)}\delta_{cb} \nnb \\
	&= \delta_{ab}\pd_{t'} \delta(t-t') - \pd_a v^c(\mb{m}(t)) \delta_{cb} \delta(t-t')\ , \label{eq:operator-O}
\end{align}
and where the notation ``$\sim$"  indicates equivalence up to infinite products of constant (but $\lam$-independent) 
factors. The final formula Eq.~\eqref{eq:EL-result} is famously equivalent to the stationary-phase approximation to $Z(T)$, but 
where the sum is taken over \emph{all} $T$-periodic solutions of the classical equations of motion, and not just the solution 
which minimizes the action. In favorable cases there are a finite number of solutions 
$\overline{\mb{m}}(t) \in  L\mathcal{M}_S$, and the partition function reduces to a sum of finitely many terms. In addition, the 
Pfaffians appearing in this expression can be computed using standard regularization techniques (see, for example, 
Ref.~\onlinecite{miettinen1996localization}), as we discuss in Appendix~\ref{app:dets} for the examples considered in this 
paper.

We note here that there is a typo in the presentation of this formula in several original references on the EL technique. The 
formula presented here is the correct one and it can be found in this form in Eq.~3.13 of Ref.~\onlinecite{keski1991topological} 
and Eq.~13 of Ref.~\onlinecite{miettinen1996localization}, for example. Note, however, that we present this formula in 
terms of an operator $\mathcal{O}$ which has all indices down, $\mathcal{O}_{ab}(t,t')$. We find that this presentation makes
more sense since typically one considers the Pfaffian of an antisymmetric bilinear form $\mathcal{O}_{ab}$ and not a linear 
operator ${\mathcal{O}^a}_b$ which happens to be antisymmetric. In addition, in the infinite-dimensional case one
needs to also properly define the Pfaffian, and with the index structure that we have chosen it is possible to define this
Pfaffian in terms of a fermion path integral as we now discuss. 

The Pfaffian of a $2n\times 2n$ antisymmetric matrix $\mathcal{O}_{ab}$ is a well-defined object, in the sense that
there is an explicit formula for it. One way of computing the Pfaffian is by Grassmann integration. 
If $\eta^a$, $a=1,\dots,2n$, are a set of $2n$ real Grassmann variables, then we have 
\beq
	\text{Pf}[\mathcal{O}]= \int d^{2n}\mbs{\eta}\ e^{-\frac{1}{2}\eta^a \mathcal{O}_{ab}\eta^b}\ ,
\eeq
provided that we define the measure as $d^{2n}\mbs{\eta}= d\eta^1\cdots d\eta^{2n}$. We therefore propose that in
the infinite-dimensional case one should define the Pfaffian of the operator $\mathcal{O}$ via the fermionic path integral
\beq
	\text{Pf}[\mathcal{O}]= \int [d^{2n}\mbs{\eta}] e^{-\frac{1}{2}\int_0^T dt \int_0^T dt' \eta^a(t) \mathcal{O}_{ab}(t,t') \eta^b(t')}\ , \label{eq:Pfaffian-as-path-integral}
\eeq
where $\eta^a(t)$ are the Grassmann-valued fields with periodic boundary conditions that we considered earlier in this 
section. We can then evaluate the integral by expanding the fields in Fourier modes as
\beq
	\eta^a(t)= \sum_{m\in\mathbb{Z}} \eta^a_m \frac{e^{i\frac{2\pi m t}{T}}}{\sqrt{T}}\ ,
\eeq
where the Fourier coefficients $\eta^a_m$ are ordinary Grassmann numbers. 
We also need to define the path integral measure. One possible definition is (we specialize to $n=1$ here)
\beq
	[d^{2}\mbs{\eta}]= d\eta^1_0 d\eta^2_0 \prod_{m>0} d\eta^1_{-m}d\eta^1_m d\eta^2_{-m} d\eta^2_m\ ,
\eeq
however, the definition of the measure is ambiguous because different orderings of the terms will lead to answers which
differ by an overall sign. This ambiguity is not important at this stage however, because we will eventually need to regulate
the result of the path integral in order to make sense of it. We consider the careful regularization of this integral for specific
examples in Sec.~\ref{sec:global} of the main text and in Appendix~\ref{app:dets}.

\section{Evaluation of Determinants}
\label{app:dets}

In this appendix we compute the amplitude and phase of the regularized determinants $\text{det}[\mathcal{D}_{\pm}]_{reg}$ 
which are needed for the calculation of the partition function $Z[A]$ for the gauged boundary theory of the BTI state in 
Sec.~\ref{sec:global} of this paper. 
We use zeta and eta functions (to be defined below) to regularize the magnitude and phase, respectively, of these 
determinants. The application of zeta and eta function methods to the regularization of determinants
appearing in the context of EL calculations was discussed in detail by Miettinen in Ref.~\onlinecite{miettinen1996localization}.
In particular, Miettinen showed that by defining the phase of the regularized
determinant using the \emph{eta invariant} of the operator in question, the character formula for $SU(2)$
(equivalent to the partition function for a spin in a constant magnetic field) could be obtained directly from an EL path integral 
calculation, without the need to correct the final answer by hand using a so-called ``Weyl shift"\footnote{For 
the spin $J$ representation of $SU(2)$, the Weyl shift refers to the replacement of $J$ with $J+\frac{1}{2}$ in the final
answer obtained from the phase space path integral.}.

In Sec.~\ref{sec:global} we showed that the expression for the determinant of $\mathcal{D}_{\pm}$ 
could be manipulated into the form
\beq
	\text{det}[\mathcal{D}_{\pm}]= \left(\prod_{m\in\mathbb{Z}} |\lam^{(\pm)}_m| \right)e^{i\frac{(2p+1)\pi}{2}\sum_{m\in\mathbb{Z}}(1-\text{sgn}(\lam^{(\pm)}_m))}\ ,
\eeq
where $p$ was an arbitrary integer. We remind the reader that $\mathcal{D}_{\pm}=  -i\pd_t \pm \overline{A}_t$,
and the eigenvalues of $\mathcal{D}_{\pm}$ are 
$\lam^{(\pm)}_m=  \frac{2\pi m}{T} \pm \overline{A}_t$, $m\in\mathbb{Z}$. In Sec.~\ref{sec:global}
we also showed that a regularization of the infinite sum $\sum_{m\in\mathbb{Z}}1$ using 
the Riemann zeta function allowed us to reduce this expression to 
\beq
	\text{det}[\mathcal{D}_{\pm}]= \left(\prod_{m\in\mathbb{Z}} |\lam^{(\pm)}_m| \right)e^{-i\frac{(2p+1)\pi}{2}\sum_{m\in\mathbb{Z}}\text{sgn}(\lam^{(\pm)}_m)}\ .
\eeq
In this appendix we show how zeta and eta function methods can be used to carefully define the amplitude and phase in
this formal expression for the determinant of $\mathcal{D}_{\pm}$.

We start with the calculation of the amplitude $\prod_{m\in\mathbb{Z}} |\lambda^{(\pm)}_m|$. To be concrete, we first assume that
$\overline{A}_t \in (0,\frac{2\pi}{T})$. In this case we have
\beq
	\prod_{m\in\mathbb{Z}} |\lambda^{(\pm)}_m|= \overline{A}_t \prod_{m>0}\left[\left(\frac{2\pi m}{T}\right)^2 -(\overline{A}_t)^2  \right]\ .
\eeq
To regularize the product on the right-hand side of this equation we first note that the ratio
\beq
	\prod_{m>0}\left[\frac{\left(\frac{2\pi m}{T}\right)^2 -(\overline{A}_t)^2}{\left(\frac{2\pi m}{T}\right)^2}  \right] = \frac{\sin\left(\frac{\overline{A}_t T}{2}\right)}{\frac{\overline{A}_t T}{2}}\ , \label{eq:ratio-dets}
\eeq
is a completely well-defined quantity. To compute this ratio we used the infinite product formula for the sine function,
\beq
	\sin(x)= x\prod_{m=1}^{\infty}\left(1-\frac{x^2}{\pi^2 m^2} \right)\ .
\eeq
The product $\prod_{m>0}\left(\frac{2\pi m}{T}\right)^2$ in the denominator on the
left-hand side of Eq.~\eqref{eq:ratio-dets} can be interpreted as $\text{det}'[-i\pd_t]$, where the prime indicates the 
determinant without the contribution from the zero mode. We can use zeta function regularization~\cite{hawking1977zeta}
to assign a finite value to this determinant. 

To apply zeta function regularization we first define a differential operator $\mathcal{P}$ with eigenvalues 
$\left(\frac{2\pi m}{T}\right)^2$, $m>0$. We then define the \emph{spectral zeta function} for this operator as
\beq
	\zeta_{\mathcal{P}}(s)= \sum_{m>0}\left(\frac{2\pi m}{T}\right)^{-2s}\ ,
\eeq
which is well-defined for $\text{Re}[s]>\frac{1}{2}$. Then the regularized version of the determinant of 
$\mathcal{P}$ is defined as
\beq
	\text{det}[\mathcal{P}]_{reg}= e^{-\zeta'_{\mathcal{P}}(0)}\ ,
\eeq
where $\zeta'_{\mathcal{P}}(0)$ is the analytic continuation of $\zeta'_{\mathcal{P}}(s)$ to $s=0$ (and the prime denotes
a derivative with respect to $s$).
In this case the spectral zeta function $\zeta_{\mathcal{P}}(s)$ is related to the ordinary Riemann zeta function $\zeta(s)$ by
\beq
	\zeta_{\mathcal{P}}(s)= \left(\frac{T}{2\pi}\right)^{2s}\zeta(2s)\ ,
\eeq
which means that
\beq
	\zeta'_{\mathcal{P}}(0) = 2\ln\left(\frac{T}{2\pi}\right)\zeta(0) + 2\zeta'(0)\ .
\eeq
Using the well-known values $\zeta(0)= -\frac{1}{2}$ and $\zeta'(0)= -\frac{1}{2}\ln(2\pi)$, we find that 
$\zeta'_{\mathcal{P}}(0)= -\ln(T)$, so that
\beq
	\text{det}[\mathcal{P}]_{reg}=T\ .
\eeq
Then, in view of the ratio Eq.~\eqref{eq:ratio-dets}, we define
\beqa
	\left(\prod_{m\in\mathbb{Z}} |\lambda^{(\pm)}_m|\right)_{reg} &=& \overline{A}_t\ \text{det}[\mathcal{P}]_{reg}\ \frac{\sin\left(\frac{\overline{A}_t T}{2}\right)}{\frac{\overline{A}_t T}{2}} \nnb \\
&=& 2 \sin\left(\frac{\overline{A}_t T}{2}\right)\ .
\eeqa
More generally, suppose that $\overline{A}_t$ lies in the open interval $(\frac{2\pi \ell}{T},\frac{2\pi \ell +2\pi}{T})$ for
some $\ell \in \mathbb{Z}$. In this case it is convenient to decompose $\overline{A}_t$ as
\beq
	\overline{A}_t = \frac{2\pi\ell}{T} + \overline{a}_t\ , \label{eq:decompose-gauge}
\eeq
where $\overline{a}_t \in (0,\frac{2\pi}{T})$. If we now repeat the amplitude calculation above for this case then we find that
\beqa
	\left(\prod_{m\in\mathbb{Z}} |\lambda^{(\pm)}_m|\right)_{reg} &=& (-1)^{\ell} 2 \sin\left(\frac{\overline{A}_t T}{2}\right) \nnb \\
	&=& (-1)^{\ell} 2 \sin\left(\pi \ell + \frac{\overline{a}_t T}{2}\right) \nnb \\
	&=& 2\sin\left(\frac{\overline{a}_t T}{2}\right)\ .
\eeqa

We now move on to the computation of the phase of the regularized determinants. First, for a complex number $s$ with 
sufficiently large and positive real part, the \emph{eta function} $\eta_{\pm}(s)$ of the Dirac operator $\mathcal{D}_{\pm}$ is 
defined by~\cite{APS1}
\beq
	\eta_{\pm}(s)= \sum_{m\in\mathbb{Z}}\text{sgn}( \lambda^{(\pm)}_m ) |\lambda^{(\pm)}_m|^{-s}\ ,
\eeq
where we use the convention that $\text{sgn}(0)=1$. This expression has a well-defined analytic continuation to $s=0$, 
known as the \emph{eta invariant}, and we use this analytic continuation to define the regularized phase of the determinant in 
question via the formula
\beq
	\left(\sum_{m\in\mathbb{Z}}\text{sgn}(\lam^{(\pm)}_m)\right)_{reg}= \eta_{\pm}(0)\ .
\eeq
We focus our attention on the calculation of the eta invariant for $\mathcal{D}_{+}$. The calculation for 
$\mathcal{D}_{-}$ is very similar. 

First, recall that we are assuming that $\overline{A}_t$ lies in an open interval between two eigenvalues
of $-i\pd_t$. This guarantees that the operators $\mathcal{D}_{\pm}$ do not possess any zero modes.
In this case each term in $\eta_{\pm}(s)$ can be differentiated with respect to $\overline{A}_t$, since the 
value of $\text{sgn}( \lambda^{(\pm)}_m )$ does not vary as we move $\overline{A}_t$ within this open interval. After
taking the derivative, we find that (focusing on the case of $\mathcal{D}_{+}$)
\beq
	\frac{d \eta_{+}(s)}{d\overline{A}_t}= -s \zeta_{\mathcal{D}_{+}^2}(\tfrac{s+1}{2})\ ,
\eeq
where $\zeta_{\mathcal{D}_{+}^2}(s)$ is the spectral zeta function for $\mathcal{D}_{+}^2$, the square of the
Dirac operator $\mathcal{D}_{+}$. This formula is in fact just a special case of the general formula in Proposition 2.10 of 
Ref.~\onlinecite{APS3}. Taking the $s\to 0$ limit then gives 
\beq
	\frac{d \eta_{+}(0)}{d\overline{A}_t}= - \lim_{s\to 0}\  s \zeta_{\mathcal{D}_{+}^2}(\tfrac{s+1}{2})\ . \label{eq:eta-derivative}
\eeq

The spectral zeta function $\zeta_{\mathcal{D}_{+}^2}(s)$ has a first order pole at $s=\frac{1}{2}$, which is due to the 
fact that the leading part of $\mathcal{D}_{+}^2$ is $-\pd_t^2$ (i.e., the dominant part of the eigenvalues of 
$\mathcal{D}_{+}^2$ for large $m$ is the piece $\left(\frac{2\pi m}{T}\right)^2$). 
It then follows from Eq.~\eqref{eq:eta-derivative} that $\frac{d \eta_{+}(0)}{d\overline{A}_t}$ is
equal to \emph{minus} the residue of $\zeta_{\mathcal{D}_{+}^2}(s)$ at $s=\frac{1}{2}$. 
This residue can be computed using the heat kernel expansion for $\mathcal{D}_{+}^2$, and the residue turns out to be equal to 
the residue of the spectral zeta function for $-\pd_t^2$ at $s=\frac{1}{2}$, which is easier to compute. From these 
considerations we find that 
\beq
	\frac{d \eta_{+}(0)}{d\overline{A}_t} = -\frac{T}{\pi}\ ,
\eeq
and then an integration with respect to $\overline{A}_t$ gives
\beq
	\eta_{+}(0)= C_{+} - \frac{\overline{A}_t T}{\pi}\ ,
\eeq
where $C_{+}$ is an as yet undetermined constant. 

The value of the constant $C_{+}$ can be fixed uniquely by requiring the eta invariant to 
vanish when $\overline{A}_t$ lies halfway between two eigenvalues of $-i\pd_t$ (symmetry dictates that $\eta_{+}(s)$ for any 
$s$ should vanish in this case). Let us assume that $\overline{A}_t \in (\frac{2\pi \ell}{T},\frac{2\pi \ell + 2\pi}{T})$ for
some $\ell\in\mathbb{Z}$. Then we require $\eta_{+}(0)$ to vanish when $\overline{A}_t=\frac{2\pi}{T}(\ell+\frac{1}{2})$, 
which fixes $C_{+}= 2\ell+1$. 
Therefore the eta invariant is given in this case by
\beq
	\eta_{+}(0)= 2\ell+ 1-\frac{\overline{A}_t T}{\pi}\ .
\eeq
For the Dirac operator $\mathcal{D}_{-}$, and still assuming that 
$\overline{A}_t \in (\frac{2\pi \ell}{T},\frac{2\pi \ell + 2\pi}{T})$, all of the signs are 
reversed. We then find that for $\overline{A}_t \in (\frac{2\pi \ell}{T},\frac{2\pi \ell + 2\pi}{T})$, the eta invariants of the 
operators $\mathcal{D}_{\pm}$ are
\beq
	\eta_{\pm}(0)= \pm (2\ell+1)\mp\frac{\overline{A}_t T}{\pi}\ .
\eeq
As in Eq.~\eqref{eq:decompose-gauge}, it is convenient to again write $\overline{A}_t= \frac{2\pi \ell}{T} + \overline{a}_t$ with 
$\overline{a}_t \in (0,\frac{2\pi}{T})$. Then in terms of $\overline{a}_t$, the eta invariants for $\mathcal{D}_{\pm}$
take the form
\beq
	\eta_{\pm}(0)= \pm 1 \mp\frac{\overline{a}_t T}{\pi}\ .
\eeq
We see that the eta invariant only depends on the value of $\overline{A}_t$ modulo $\frac{2\pi}{T}$.


%

\end{document}